\documentclass[11pt,a4paper]{article}
\pdfoutput=1
\usepackage{jheppub}
\allowdisplaybreaks
\usepackage{graphicx,psfrag}
\usepackage{bm}
\usepackage{mathbbol,verbatim}
\usepackage{slashed}
\usepackage{lscape}
\usepackage{graphics}
\usepackage{color,ulem}
\usepackage{multirow}
\graphicspath{{figs/}}

\allowdisplaybreaks

\definecolor{greeen}{rgb}{0.03,0.84,0.13}
\definecolor{test}{rgb}{0.03,0.74,0.33}
\definecolor{viol}{rgb}{0.44,0,0.94}
\definecolor{or}{rgb}{0.95,0.65,0}

\newcommand{\tabincell}[2]{\begin{tabular}{@{}#1@{}}#2\end{tabular}}

\definecolor{greeen}{rgb}{0.03,0.84,0.13}
\definecolor{test}{rgb}{0.03,0.74,0.33}
\definecolor{viol}{rgb}{0.44,0,0.94}
\definecolor{or}{rgb}{0.95,0.65,0}

\title{Leptonic Scalars at the LHC }

\author[a]{Andr\'{e} de Gouv\^{e}a,}
\author[b]{P. S. Bhupal Dev,}
\author[c]{Bhaskar Dutta,}
\author[d]{Tathagata Ghosh,}
\author[e]{Tao Han,}
\author[b]{Yongchao Zhang}

\affiliation[a]{Department of Physics and Astronomy, Northwestern University, Evanston, IL 60208, USA}
\affiliation[b]{Department of Physics and McDonnell Center for the Space Sciences,  Washington University, St. Louis, MO 63130, USA}
\affiliation[c]{Mitchell Institute for Fundamental Physics and Astronomy, Department of Physics and Astronomy, Texas A\&M University, College Station, Texas 77843, USA}
\affiliation[d]{Department of Physics and Astronomy, University of Hawaii, Honolulu, HI 96822, USA}
\affiliation[e]{Pittsburgh Particle Physics, Astrophysics, and Cosmology Center, Department of Physics and Astronomy, University of Pittsburgh,
3941 O'Hara St., Pittsburgh, PA 15260, USA}

\keywords{Neutrino Self-interactions, Leptonic Scalars, Large Hadron Collider}

\preprint{~~PITT-PACC 1909, MI-TH-1936}

\date{\today}

\begin{document}

\abstract{We explore the collider prospects of neutrino non-standard interaction with a Standard Model (SM) gauge-singlet leptonic scalar $\phi$ carrying two units of lepton-number-charge. These leptonic scalars are forbidden from interacting with the SM fermions at the renormalizable level and, if one allows for higher-dimensional operators, couple predominantly to SM neutrinos. For masses at or below the electroweak scale, $\phi$ decays exclusively into neutrinos. Its characteristic production signature at hadron collider experiments like the LHC would be via the vector boson fusion process and leads to same-sign dileptons, two forward jets in opposite hemispheres, and missing transverse energy, i.e., $pp \to \ell_\alpha^\pm \ell_\beta^\pm jj + E_T^{\rm miss}$ ($\alpha,\, \beta = e,\, \mu, \tau$). Exploiting the final states of electrons and muons, we estimate, for the first time, the sensitivity of the LHC to these lepton-number-charged scalars. We show that the LHC sensitivity is largely complementary to that of low-energy precision measurements of the decays of charged leptons, charged mesons, $W$, $Z$ and the SM Higgs boson, as well as the neutrino beam experiments like MINOS, and searches for  neutrino self-interactions at IceCube and in cosmological observations. For $\phi$ mass larger than roughly 10 GeV, our projected LHC sensitivity would surpass all existing bounds.
}

\maketitle

%

\section{Introduction}

Over the past two decades, neutrino oscillations have been observed in the solar, atmospheric, reactor and accelerator neutrino experiments, revealing that at least two of the three neutrinos in the Standard Model (SM) are massive particles \cite{PDG}. Yet, neutrinos remain most elusive and many questions in the neutrino sector need to be answered. Those include: (i) Are neutrinos Dirac-type or Majorana-type fermions? (ii) Is the lightest neutrino predominantly coupled to electrons in charged-current weak interactions? (iii) Is CP-invariance violated in the lepton sector? (iv) Are there non-standard interactions involving neutrinos that go beyond their mass generation? To answer these outstanding questions, the study of neutrino properties at all accessible experiments is strongly motivated.

If neutrinos are massive Dirac fermions, lepton-number (or some non-anomalous symmetry that contains lepton-number, such as $B-L$) is a conserved symmetry in nature. In this case, new, hypothetical particles can be characterized according to their lepton-number-charge and states associated to different lepton-number-charge will behave qualitatively differently~\cite{Rao:1983sd, deGouvea:2014lva, Kobach:2016ami}. For example, new scalars with lepton-number-charge equal to one only couple in pairs to SM particles and are interesting dark matter (DM) candidates \cite{Berryman:2018ogk,Kelly:2019wow}. On the other hand, a new scalar with lepton-number-charge equal to minus two, denoted by $\phi$ and henceforth dubbed as a ``leptonic scalar'', can only couple individually to right-handed neutrinos ($\nu^c$) like $\nu^c\nu^c\phi^*$ at the renormalizable level. At the dimension-six level, it also couples to a pair of lepton-doublets ($L$) and Higgs-doublets ($H$) like  $(LH)(LH)\phi/\Lambda^2$, where $\Lambda$ is the new physics scale that gives rise to this dimension-six operator. After electroweak (EW) symmetry breaking, the latter yields the low-energy effective Lagrangian
\begin{eqnarray}
\label{eqn:Lagrangian}
{\cal L} \ \supset \
\frac12 \lambda_{\alpha\beta}\ \phi\ \nu_\alpha \nu_\beta \,,
\end{eqnarray}
where $\alpha,\, \beta = e,\, \mu,\, \tau$ are the lepton-flavor indices and $\lambda_{\alpha\beta}$ the flavor-dependent Yukawa couplings of order $v^2/\Lambda^2$, with $v\equiv  (\sqrt2 G_F)^{-1/2} \simeq 246$ GeV being the EW scale (with $G_F$ being the Fermi constant). To be self-consistent, within the effective field theory (EFT) framework, we concentrate on scalar masses $m_\phi<v$. Examples of concrete ultraviolet (UV)-complete models that could give rise to the effective Lagrangian~\eqref{eqn:Lagrangian} below the EW scale are discussed in Appendix~\ref{app:UV}. Note that the couplings in Eq.~(\ref{eqn:Lagrangian}) define one class of well-motivated simplified models for non-standard neutrino self-interactions (see Ref.~\cite{Dev:2019anc} for a recent review); if the momentum transfer is much smaller than the scalar mass $m_\phi$, then the scalar $\phi$ can be integrated out and we are left with the effective four-neutrino interactions~\cite{Blinov:2019gcj}.

We should mention that the results discussed below will also apply if the neutrinos are Majorana fermions under the assumption that lepton-number violating effects are very small and effectively absent at collider experiments. For example, if very heavy Majorana masses $M_{\nu^c}\gg v$ for the right-handed neutrinos are added to the SM Lagrangian (along with the neutrino Yukawa couplings), which make up the only source of lepton-number violation, then lepton-number symmetry is approximately conserved at collider energies. In this case, it is fair to assign a lepton-number-charge to $\phi$ and assume that its main coupling to the SM is via the dimension-six operator of interest.

\begin{figure}
  \centering
  \includegraphics[height=0.35\textwidth]{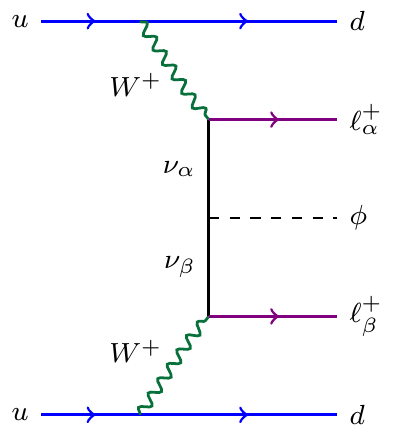}
  \caption{Representative Feynman diagram for the production of leptonic scalar $\phi$ at the LHC.}
  \label{fig:diagram}
\end{figure}

Given the interaction Lagrangian~\eqref{eqn:Lagrangian}, the leptonic scalar $\phi$ can be produced by radiation off a neutrino. As such, there is a large class of processes at different energy regime to search for its existence, as we will discuss in detail. In particular, at high-energy hadron colliders, it can be produced in a characteristic sub-process like
\begin{eqnarray}
\label{eqn:signal}
uu \ \to \ dd\ \ell^+_\alpha \ell^+_\beta\ \phi \,,
\end{eqnarray}
where $\phi$ decays subsequently into neutrinos and hence manifests itself as missing energy in the vector-boson fusion (VBF) process. Generically, $\phi$-production is characterized by same-sign dileptons plus two forward jets and missing transverse energy. The corresponding Feynman diagram is depicted in Fig.~\ref{fig:diagram}.   This topology is the same as the one for the emission of a Majoron from neutrinoless double beta  ($0\nu\beta\beta$) decay process~\cite{Georgi:1981pg, Doi:1985dx}. For Majoron masses smaller than ${\cal O}$(MeV) -- the typical $Q$-value for relevant nuclei, strong limits on the coupling {$\lambda_{ee}\lesssim 10^{-4}$}~\cite{Berryman:2018ogk} have been set by $0\nu\beta\beta$ experiments like NEMO-3~\cite{Arnold:2013dha, Arnold:2015wpy, Arnold:2016ezh, Arnold:2016qyg, Arnold:2018tmo, NEMO-3:2019gwo}, KamLAND-Zen~\cite{Gando:2012pj},  EXO-200~\cite{Albert:2014fya} and GERDA~\cite{Agostini:2015nwa}. In this paper, we show that high-energy colliders like LHC provide a novel complementary probe of the couplings $\lambda_{\alpha\beta}$ through the VBF process~\eqref{eqn:signal} that extends the experimental reach to  relatively higher $\phi$ masses. Note that if neutrinos were Majorana particles, one could have the lepton-number-violating  process $pp \to \ell^\pm \ell^\pm jj$ at high-energy colliders, either via the VBF channel shown in Fig.~\ref{fig:diagram} without the $\phi$ emission, or via the $s$-channel Keung-Senjanovi\'{c} process~\cite{Keung:1983uu} involving heavy Majorana neutrinos (and heavy gauge bosons). For reviews on the current constraints and future prospects of these  lepton-number-violating processes at colliders, as well as other relevant low-energy searches, including meson decays and beam dump experiments; see e.g., Refs.~\cite{Atre:2009rg, Deppisch:2015qwa, deGouvea:2015euy, Cai:2017mow, Das:2018hph}.
The process under consideration in Eq.~(\ref{eqn:signal}) has an additional leptonic scalar $\phi$ that carries away missing energy and lepton-number. We would like to clarify here that although this process by itself may not uniquely distinguish any specific UV-complete model, such as those discussed in Appendix~\ref{app:UV}, it can be used in conjunction with additional signals arising in specific UV-completions to probe the leptonic scalar at the LHC.

In this paper, we explore the impact of the couplings $\lambda_{\alpha\beta}$, defined in Eq.~(\ref{eqn:Lagrangian}), at the $\sqrt s=14$ TeV LHC with an integrated luminosity of 300 fb$^{-1}$ and the high-luminosity  upgrade (HL-LHC), up to an integrated luminosity of 3 ab$^{-1}$, as a function of $m_\phi$. We find that the LHC (HL-LHC) is sensitive to couplings $\lambda_{\alpha\beta}$ as small as $1.00$ ($0.68$) for $m_\phi \lesssim 50$~GeV. The sensitivity degrades slowly for larger  $m_\phi$, as the production cross section becomes smaller.  The LHC prospects already exceed all the current existing limits for
$m_\phi \gtrsim 10$ GeV while limits from lepton and meson decays and other low-energy data are more stringent for smaller $m_\phi$~\cite{Lessa:2007up, Pasquini:2015fjv, Berryman:2018ogk}. Hence, searches for $\phi$ at the high-energy colliders are largely complementary to those at low-energy, high-precision setups. With higher energies and larger luminosities, the sensitivity to $\lambda_{\alpha\beta}$ is expected to be improved at the $\sqrt s=27$ TeV High-Energy LHC ~\cite{Zimmermann:2018wdi} and future 100 TeV colliders like Future Circular Collider (FCC-hh)~\cite{Benedikt:2018csr} and Super Proton-Proton Collider (SPPC)~\cite{CEPC-SPPCStudyGroup:2015csa}. Studies associated to these future machines, however, go beyond the main scope of this paper and will be pursued elsewhere.

The rest of this paper is organized as follows: All the current low-energy limits on the mass $m_\phi$ and couplings $\lambda_{\alpha\beta}$ are collected in section~\ref{sec:limits}. Our estimates for the sensitivity at  the LHC and HL-LHC to the new couplings $\lambda_{\alpha\beta}$ are given in section~\ref{sec:LHC}. We present our conclusions in section~\ref{sec:conclusion}. Possible  (UV-)completions of the effective Lagrangian~\eqref{eqn:Lagrangian} are discussed in Appendix~\ref{app:UV}. Some details of the computation of the multi-body decays involving $\phi$ are relegated to Appendix~\ref{sec:decay}.

\section{Low-energy Constraints}
\label{sec:limits}

The scalar mass $m_\phi$ and the couplings $\lambda_{\alpha\beta}$ are constrained by a variety of high-precision data at low energy~\cite{Lessa:2007up, Pasquini:2015fjv, Berryman:2018ogk}.
In this section, we focus mainly on the constraints for $m_\phi > 100$ MeV, including decay rates of tauon and charged mesons, the searches of heavy neutrinos from charged meson decays, the invisible decay width of $Z$ boson and SM Higgs boson $h$, the production and decays of $W$ boson at colliders, neutrino-matter scatting in neutrino beam experiments MINOS and DUNE,
and the IceCube and cosmic microwave background (CMB) limits on the new neutrino--neutrino interactions. All of these limits are collected in Table~\ref{tab:limits} and detailed in the following subsections \ref{sec:meson}--\ref{sec:ic}.
The light scalar $\phi$ can in principle be produced in the  high-intensity beam-dump experiments like NA64 and LDMX, but the sensitivities are highly suppressed, as discussed in subsection~\ref{sec:others}.
There are also many other limits which are relevant for a lighter $\phi$ with mass $m_\phi < 100$ MeV, such as those from muon decays, tritium decay, searches of Majoron in $0\nu\beta\beta$ decay experiments, supernova, relativistic degrees of freedom $\Delta N_{\rm eff}$ in the early Universe, and the neutrino decay constraints. To be complete, all of these are summarized in subsection~\ref{sec:others}, but not used in our analysis, mainly because the new LHC sensitivities derived here become competitive only in the high-mass regime with $m_\phi\gtrsim 100$ MeV. In section~\ref{sec:LHC}, we focus only on the LHC prospects for the couplings $\lambda_{ee,\, e\mu,\, \mu\mu}$ that do not involve the $\tau$-lepton flavor in the final state shown in Fig.~\ref{fig:diagram}, therefore we exclude the couplings $\lambda_{\alpha\beta}$ involving the $\tau$ flavor from Table~\ref{tab:limits} and Figs.~\ref{fig:meson}--\ref{fig:Zinv}. For completeness, we will comment on the limits on $\tau$-flavor relevant couplings in the text, when they are applicable.

\begin{table}[!t]
  \centering
  \caption{\label{tab:limits} Summary of current and future experimental data which can be used to set limits on the couplings $|\lambda_{\alpha\beta}|$ (with $\ell = e,\,\mu$) or their combinations. The last column shows the relevant $m_\phi$ ranges (see Figs.~\ref{fig:meson}--\ref{fig:Zinv} and \ref{fig:LHC}--\ref{fig:LHC3}). For the limits from invisible $Z$ and $h$ decays, the symmetry factor $S_{\alpha\beta} = 1 \, (1/2)$ for $\alpha \neq \beta$ ($\alpha = \beta$). For the invisible Higgs decay and  IceCube data, the numbers in the parentheses are respectively the expected sensitivity at the HL-LHC and IceCube-Gen2. The limits not collected in this table are either weaker or not relevant for $m_\phi > 100$ MeV. The branching fraction  (BR) upper limits  are at 95\% confidence level (C.L.), whereas the error bars quoted for the BR measurements are at $1\sigma$ C.L.; see text for more details. }
  \vspace{0.2cm}
  \scriptsize
  \begin{tabular}{rllcc}
  \hline\hline
  {\bf Ref.} & {\bf Process} &  {\bf Data} & {\bf Couplings} & {\bf Mass range} \\ \hline\hline
  \cite{PDG, Berryman:2018ogk} & $\pi^- \to e^- \bar\nu_e\nu\bar\nu$ &  ${\rm BR} < 5\times 10^{-6}$ & $\sum_\beta |\lambda_{e\beta}|^2$ & $m_\phi < 131$ MeV \\ \hline

  \cite{PDG, Berryman:2018ogk} & $K^- \to e^- \bar\nu_e\nu\bar\nu$   &  ${\rm BR} < 6\times 10^{-5}$ & $\sum_\beta |\lambda_{e\beta}|^2$ & $m_\phi < 444$ MeV \\
  \cite{PDG, Berryman:2018ogk} & $K^- \to \mu^- \bar\nu_\mu \nu\bar\nu$   &  ${\rm BR} < 2.4\times 10^{-6}$ & $\sum_\beta |\lambda_{\mu\beta}|^2$ & $m_\phi < 386$ MeV \\ \hline

  \cite{PDG, Berryman:2018ogk} & $D^- \to e^- \bar\nu_e$   & ${\rm BR} < 8.8\times 10^{-6}$ & $\sum_\beta |\lambda_{e\beta}|^2$ & $m_\phi < 1.52$ GeV \\
  \cite{PDG, Berryman:2018ogk} & $D^- \to \mu^- \bar\nu_\mu$  &   ${\rm BR} < 3.4\times 10^{-5}$ & $\sum_\beta |\lambda_{\mu\beta}|^2$ & $m_\phi < 1.39$ GeV \\ \hline

  \cite{PDG, Pasquini:2015fjv} & $D_s^- \to e^- \bar\nu_e$  &   ${\rm BR} < 8.3\times 10^{-5}$ & $\sum_\beta |\lambda_{e\beta}|^2$ & $m_\phi < 1.64$ GeV \\
  \cite{PDG, Pasquini:2015fjv} & $D_s^- \to \mu^- \bar\nu_\mu$  &   ${\rm BR} = (5.50 \pm 0.23)\times 10^{-3}$ & $\sum_\beta |\lambda_{\mu\beta}|^2$ & $m_\phi < 1.50$ GeV \\ \hline

  \cite{PDG, Pasquini:2015fjv} & $B^- \to e^-\bar\nu_e$  &   ${\rm BR} < 9.8\times 10^{-7}$ & $\sum_\beta |\lambda_{e\beta}|^2$ & $m_\phi < 3.54$ GeV \\
  \cite{PDG, Pasquini:2015fjv} & $B^- \to \mu^- \bar\nu_\mu$  &   ${\rm BR} = (2.90 - 10.7) \times 10^{-7}$ & $\sum_\beta |\lambda_{\mu\beta}|^2$ & $m_\phi < 3.50$ GeV \\ \hline

  \cite{PDG, Lessa:2007up} & $\tau^- \to e^- \bar\nu_e \nu_\tau$ & ${\rm BR} = (17.82 \pm 0.04) \%$ & $\sum_\beta |\lambda_{e\beta}|^2$ & $m_\phi < 741$ MeV \\
  \cite{PDG, Lessa:2007up} & $\tau^- \to \mu^- \bar\nu_\mu \nu_\tau$ & ${\rm BR} = (17.39 \pm 0.04) \%$ & $\sum_\beta |\lambda_{\mu\beta}|^2$ & $m_\phi < 741$ MeV \\ \hline

  \cite{PDG, Pasquini:2015fjv} & $P^- \to e^- N$ & see Ref.~\cite{CortinaGil:2017mqf} & $\sum_\beta |\lambda_{e\beta}|^2$ & $3.3 \, {\rm MeV} < m_\phi < 448$ MeV \\
  \cite{PDG, Pasquini:2015fjv} & $P^- \to \mu^- N$ & see Ref.~\cite{CortinaGil:2017mqf} & $\sum_\beta |\lambda_{\mu\beta}|^2$ & $87 \, {\rm MeV} < m_\phi < 379$ MeV \\ \hline


  \multirow{2}{*}{\cite{PDG}}  & \multirow{2}{*}{$Z \to {\rm inv.}$}  &
  $\Gamma_{\rm obs}^{\rm inv} = (499.0 \pm 1.5)$ MeV & \multirow{2}{*}{$\sum_{\alpha,\,\beta} S_{\alpha\beta} |\lambda_{\alpha\beta}|^2$} & \multirow{2}{*}{$m_\phi < 52.2$ GeV} \\
  & & $\Gamma_{\rm SM}^{\rm inv} = (501.44 \pm 0.04)$ MeV &  & \\  \hline

  \cite{PDG} & $W \to e\nu$ & ${\rm BR} = (10.71 \pm 0.16)\%$ & $\sum_\beta |\lambda_{e\beta}|^2$ & $m_\phi < 38.8$ GeV \\
  \cite{PDG} & $W \to \mu\nu$ & ${\rm BR} = (10.63 \pm 0.15)\%$ & $\sum_\beta |\lambda_{\mu\beta}|^2$ & $m_\phi < 39.3$ GeV \\ \hline

  \cite{Khachatryan:2016whc, Cepeda:2019klc} & $h \to {\rm inv.}$ &  ${\rm BR} < 24\% \, (4.2\%)$ & $\sum_{\alpha,\,\beta} S_{\alpha\beta} |\lambda_{\alpha\beta}|^2$ & $m_\phi < 64.8 \, (72.6)$ GeV \\ \hline

  \cite{Berryman:2018ogk} & MINOS 
  & see Ref.~\cite{Berryman:2018ogk} & $|\lambda_{\mu\mu}|$ & $m_\phi < 1.67$ GeV \\
  \cite{Berryman:2018ogk} & DUNE 
  & see Ref.~\cite{Berryman:2018ogk} & $|\lambda_{\mu\mu}|$ & $m_\phi < 3.00$ GeV \\ \hline


  \cite{Ng:2014pca, Ioka:2014kca} & IceCube & see Ref.~\cite{Ioka:2014kca} & $|\lambda_{\alpha\beta}|$ & $m_\phi < 2.0 \, (15.0)$ GeV \\ \hline\hline
  \end{tabular}
\end{table}

\subsection{Meson decay rates}
\label{sec:meson}

\begin{figure}
  \centering
  \includegraphics[height=0.35\textwidth]{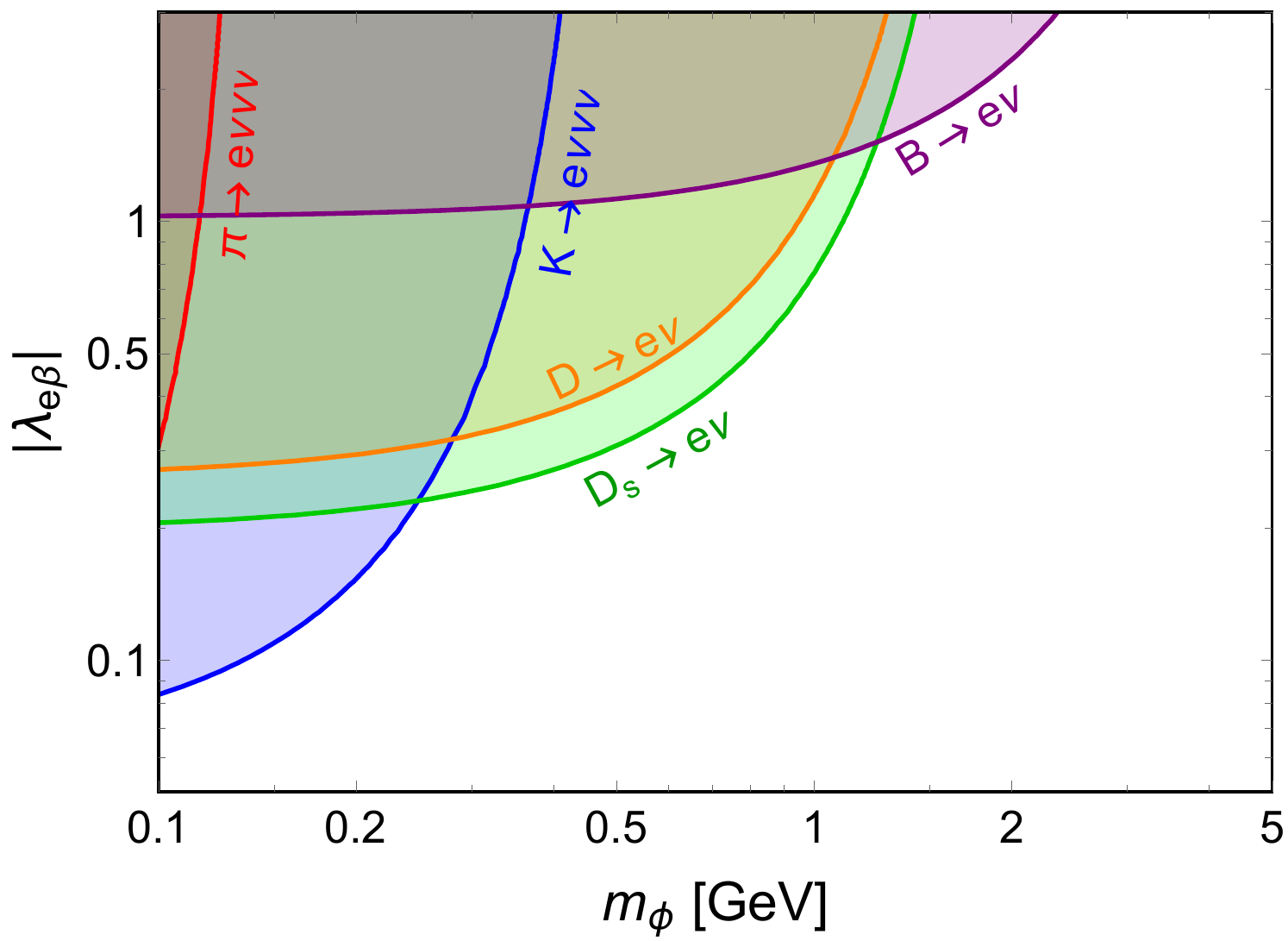}
  \includegraphics[height=0.35\textwidth]{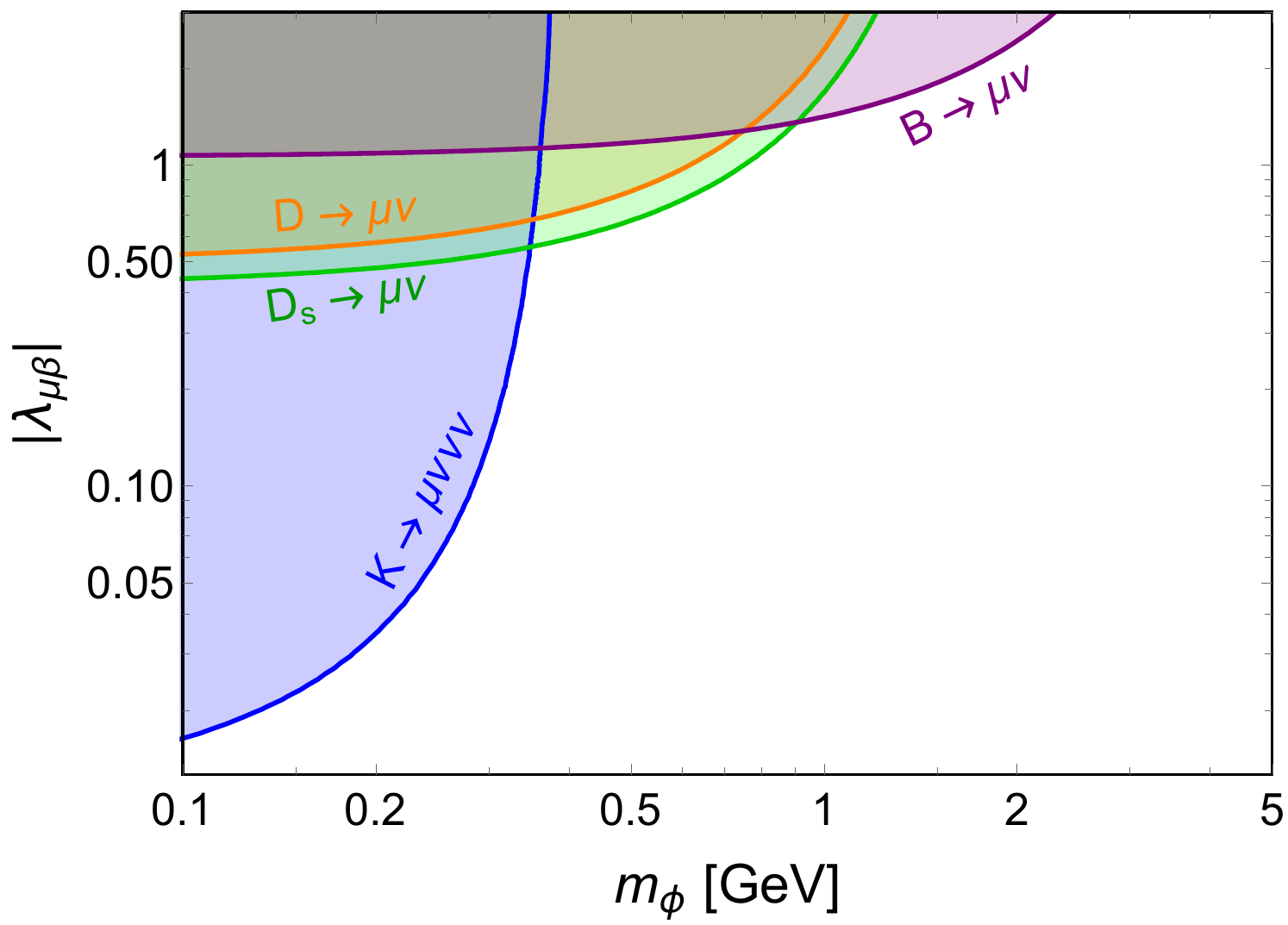}
  \caption{Limits on $|\lambda_{e\beta}|$ (left panel) and $|\lambda_{\mu\beta}|$ (right panel) with $\beta = e,\,\mu,\,\tau$ from current-charged meson decay data in Table~\ref{tab:limits}. All the shaded regions are excluded.}
  \label{fig:meson}
\end{figure}

For leptonic decays of charged mesons $P^- \to \ell^- \bar\nu$ with $P^- = \pi^-,\, K^-,\, D^-,\, D_S^-,\, B^-$, the leptonic  scalar $\phi$ can be emitted from the neutrino line in the final state and this process is not suppressed by the helicity of the charged lepton, with the partial width~\cite{Pasquini:2015fjv}
\begin{eqnarray}
\Gamma (P^- \to \ell_\alpha^- \bar\nu \phi) & \ = \ &
\frac{G_F^2 |V_{qq^\prime}|^2 m_P^3 f_P^2 \sum_\beta |\lambda_{\alpha\beta}|^2}{256 \pi^3} \nonumber \\
&&  \!\!\!\!  \!\!\!\!  \!\!\!\!  \times \int_{x_\phi}^{(1-\sqrt{x_\ell})^2} {\rm d}x
\frac{\left( (x+x_\ell) - (x-x_\ell)^2 \right)
 (x-x_\phi)^2}{x^3} \lambda^{1/2}(1,\,x,\,x_\ell) \,,
\end{eqnarray}
where
$\lambda (a,\,b,\,c) \ \equiv \ a^2 + b^2 + c^2 - 2ab - 2ac -2bc$,
$V_{qq^\prime}$ are the CKM matrix elements with the valence quarks $q$ and $q^\prime$ for the meson, $m_P$ and $f_P$ are respectively the meson mass and decay constant, $x_{\ell,\,\phi} \equiv m_{\ell,\,\phi}^2/m_P$, with $m_{\ell}$ the charged lepton mass, the charged lepton flavor $\alpha = e,\, \mu$ and we have summed over the neutrino flavor $\beta = e,\, \mu,\, \tau$ in the final state. For the light mesons $\pi^\pm$ and $K^\pm$, the rare decays  $\pi^-,\, K^- \to e^- \nu_e \nu \bar\nu$ and $K^- \to \mu^- \nu_\mu \nu \bar\nu$ have been searched for in experiments~\cite{Picciotto:1987pp, Heintze:1977kk, Artamonov:2016wby}, and the upper limits on the branching fractions are collected in Table~\ref{tab:limits}. These processes correspond to the decays $\pi^-,\, K^- \to \ell^- \bar\nu \phi$ with $\phi \to \nu\nu$, and the experimental data can be used to set limits on the couplings $\sum_\beta |\lambda_{e\beta}|^2$ and $\sum_\beta |\lambda_{\mu\beta}|^2$. For simplicity, we assume only one of the couplings $\lambda_{\alpha\beta}$ to be non-vanishing while deriving the limits. The results for $|\lambda_{e\beta}|$ and $|\lambda_{\mu\beta}|$ are shown respectively in the left and right panels of Fig.~\ref{fig:meson}.
For the heavier mesons $D^\pm$, $D_s^\pm$ and $B^\pm$, we adopt the experimental upper limits  on the BRs at the 95\% C.L. in Table~\ref{tab:limits}~\cite{PDG} to set limits on the couplings $|\lambda_{\alpha\beta}|$, as shown in Fig.~\ref{fig:meson}. For the measurements with $1\sigma$ error bars in Table~\ref{tab:limits}, we also obtain the 95\% C.L. limits on the $\lambda_{\alpha\beta}$ couplings by simply multiplying the error bars by a factor of 1.96. There are also some limits on the meson decays to tauon leptons~\cite{PDG}; however, these limits are too weak to impose any constraints on the couplings $\lambda_{\tau\beta}$.

\subsection{Heavy neutrino searches in meson decay spectra}

\begin{figure}[t!]
  \centering
  \includegraphics[height=0.35\textwidth]{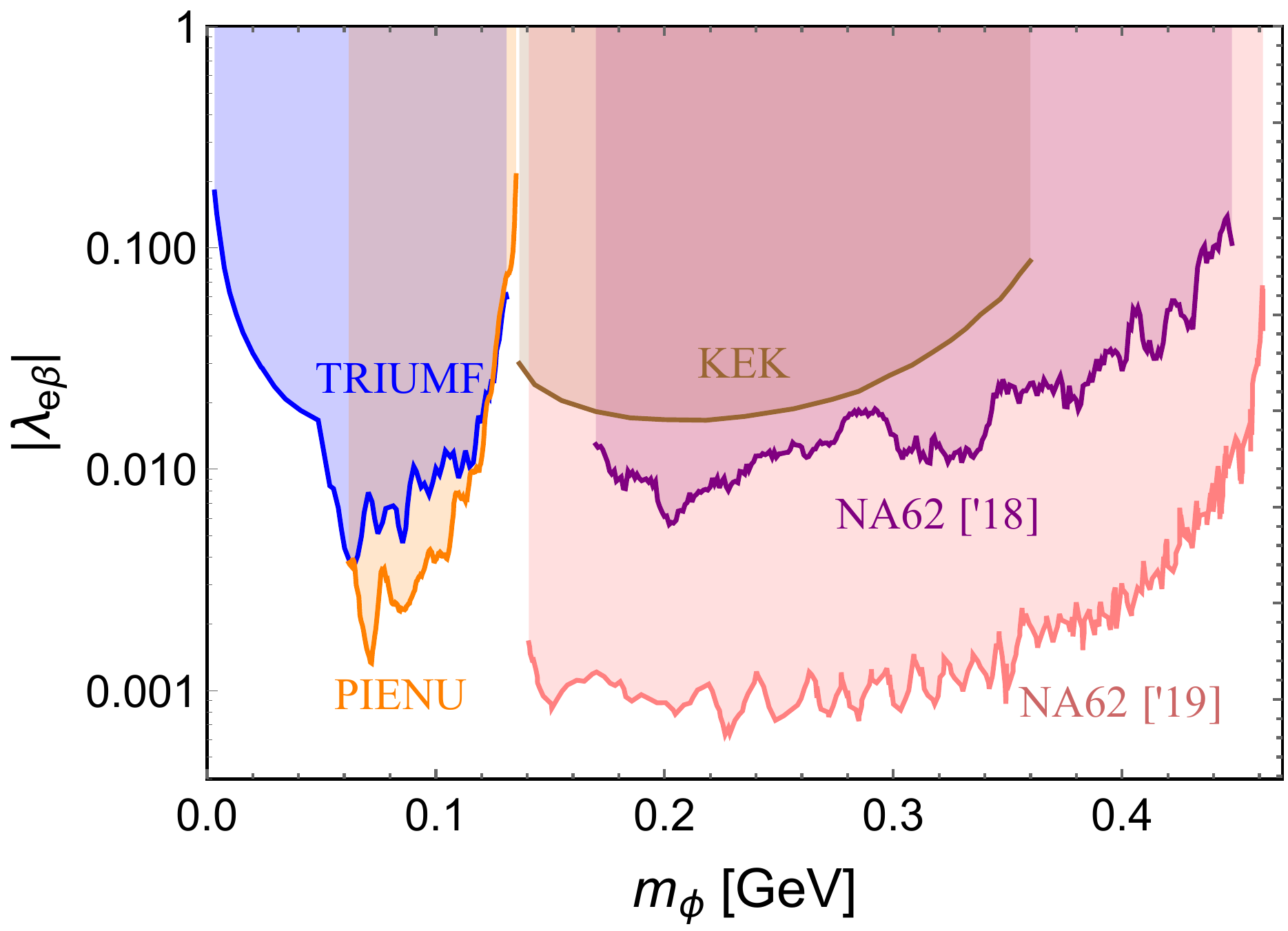}
  \includegraphics[height=0.35\textwidth]{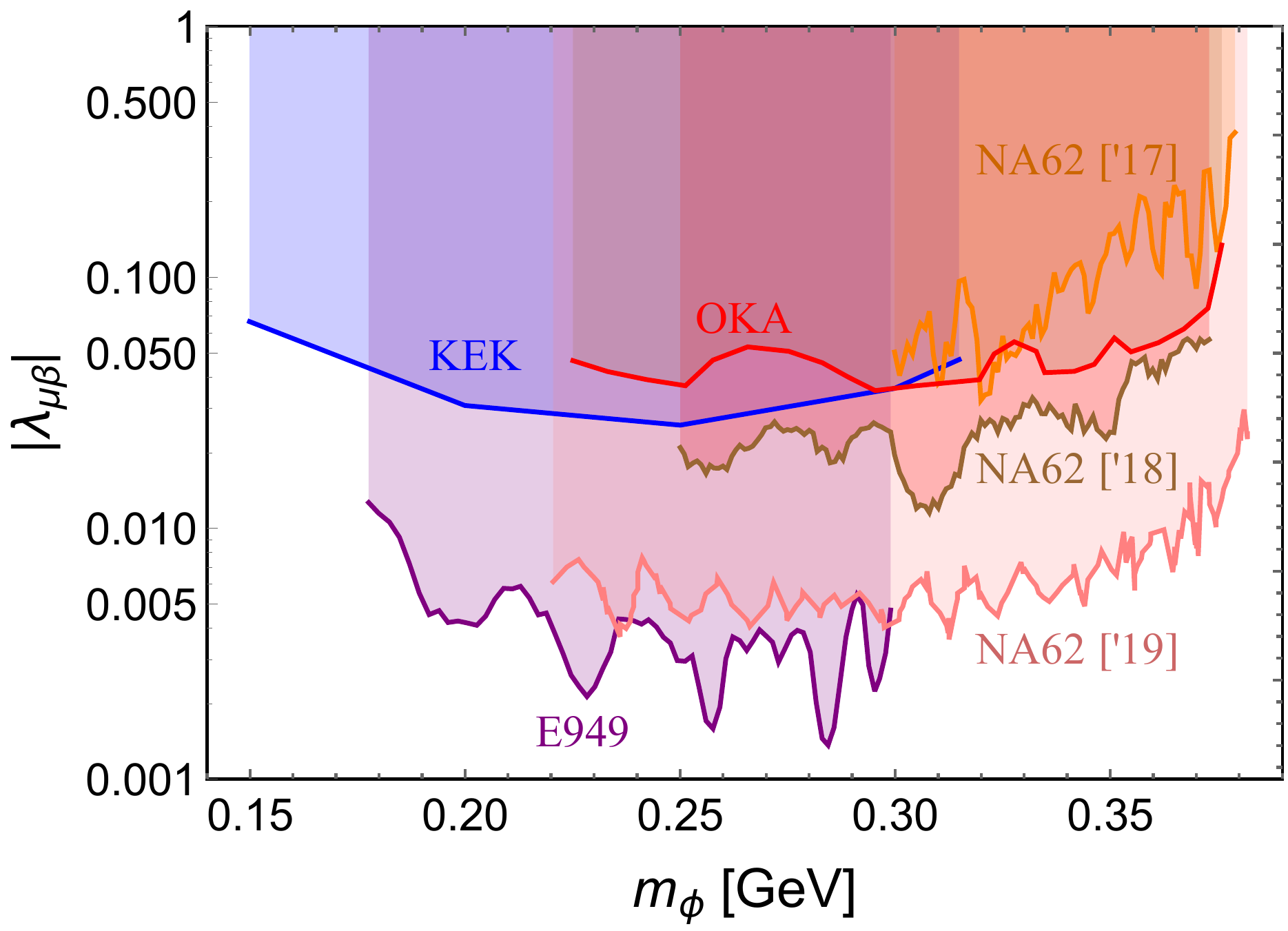}
  \caption{Limits on $|\lambda_{e\beta}|$ (left panel) and $|\lambda_{\mu\beta}|$ (right panel) with $\beta = e,\, \mu,\, \tau$ from heavy neutrino searches in meson decays in TRIUMF~\cite{Britton:1992xv}, PIENU~\cite{Aguilar-Arevalo:2017vlf}, KEK~\cite{KEK}, E949~\cite{Artamonov:2014urb},  OKA~\cite{Sadovsky:2017qsr}, NA62 ['17]~\cite{Lazzeroni:2017fza}, NA62 ['18]~\cite{CortinaGil:2017mqf} and NA62 ['19]~\cite{NA:2019}. The shaded regions are excluded.}
  \label{fig:meson2}
\end{figure}

Heavy neutrinos $N$ have been searched for in two-body meson decays, such as $\pi^- \to e^- N$ and $K^- \to \ell^- N$ (with $\ell = e,\,\mu$) by several experiments, including TRIUMF~\cite{Britton:1992xv}, PIENU~\cite{Aguilar-Arevalo:2017vlf}, KEK~\cite{KEK}, E949~\cite{Artamonov:2014urb}, OKA~\cite{Sadovsky:2017qsr} and NA62~\cite{Lazzeroni:2017fza, CortinaGil:2017mqf, NA:2019}. The peak searches in the lepton energy spectrum can be used to set limits on the leptonic scalar couplings to neutrinos, by comparing the lepton spectra of the two-body decays $P^- \to \ell^- N$ to those of the three-body decays $P^- \to \ell \nu \phi$~\cite{Pasquini:2015fjv}. For the two-body decays of charged mesons, the differential partial width with respect to the charged lepton momentum $p_\ell$ is given by~\cite{Pasquini:2015fjv}
\begin{eqnarray}
\frac{{\rm d}}{{\rm d}p_\ell} \Gamma (P^- \to \ell^- N) \ \simeq \ \rho \Gamma_0 (P^- \to \ell^- \bar\nu) |U_{\ell N}|^2
\delta (p_{\rm peak} - p_\ell)
\end{eqnarray}
with the peak position $p_{\rm peak} = \lambda^{1/2} (m_P,\, m_\ell,\, m_N)/2m_P$, and $\Gamma_0 (P^- \to \ell^- \nu)$ the leading order (LO) leptonic meson decay width in the SM:
\begin{eqnarray}
\Gamma_0 (P^- \to \ell_\alpha^- \bar\nu) & \ = \ &
\frac{G_F^2 |V_{qq^\prime}|^2 m_P^3 f_P^2}{8 \pi}
x_\ell (1-x_\ell)^2 \,,
\end{eqnarray}
$|U_{\ell N}|$ is the heavy-light neutrino mixing angle, and
\begin{eqnarray}
\rho \ = \ \frac{x_\ell+x_N - (x_\ell-x_N)^2}{x_\ell (1-x_\ell)^2} \lambda^{1/2} (1,\, x_\ell,\, x_N) \,,
\end{eqnarray}
where we have defined $x_N \equiv m_{N}^2/m_P^2$ with $m_{N}$ being the heavy neutrino mass. For the three-body decays, on the other hand,
\begin{eqnarray}
\frac{{\rm d}}{{\rm d}p_\ell}
\Gamma (P^- \to \ell_\alpha^- \nu \phi) & \ = \ &
\frac{G_F^2 |V_{qq^\prime}|^2 m_P^3 f_P^2 |\lambda_{\alpha\beta}|^2}{128\pi^3}
\left[ (x+x_\ell) - (x - x_\ell)^2 \right] \nonumber \\
&& \times \frac{(x-x_\phi)^2}{x^3 \sqrt{x_\ell^2 + p_\ell^2/m_P^2}}
\frac{p_\ell}{m_P^2}
\lambda^{1/2} (1,x,x_\ell) \,,
\end{eqnarray}
where $x \equiv 1 + x_\ell - 2 \sqrt{x_\ell + p_\ell^2/m_P^2}$. By setting $m_\phi$ equal to $m_{N}$, and demanding that the lepton energy spectrum in the three-body case should not exceed the expected spectrum for the two-body case at peak~\cite{Pasquini:2015fjv}, the resultant limits on the couplings $|\lambda_{e\beta}|$ and $|\lambda_{\mu\beta}|$ are collected respectively in the left and right panels of Fig.~\ref{fig:meson2}. As the constraints on heavy-light neutrino mixing angle $|U_{\ell N}|$ are very stringent, the limits on $|\lambda_{\alpha\beta}|$ from the meson decay spectra are very strong, down to $\sim 10^{-3}$. 

In addition to the two-body decays of meson, heavy neutrino can also be searched for by (partially) reconstructing the decay products of heavy neutrino, for instance, in the decay chain $B \to X \ell N$, $N \to \ell\pi,\, \ell_\alpha^+\ell_\beta^- \nu$ with $X$ being any SM particle. Such direct searches have been performed in the experiments PS191~\cite{Bernardi:1985ny, Bernardi:1987ek}, BEBC~\cite{CooperSarkar:1985nh}, NA3~\cite{Badier:1986xz}, NuTeV~\cite{Vaitaitis:1999wq}, LHCb~\cite{Aaij:2012zr,Aaij:2014aba,Shuve:2016muy}, Belle~\cite{Liventsev:2013zz}, NA48/2~\cite{CERNNA48/2:2016tdo} and T2K~\cite{Abe:2019kgx}, and also proposed in future experiments like FASER~\cite{Kling:2018wct, Ariga:2018uku}, MATHUSLA~\cite{Chou:2016lxi, Curtin:2018mvb}, CODEX-b~\cite{Gligorov:2017nwh, Helo:2018qej}, AL3X~\cite{Dercks:2018wum},  SHiP~\cite{Alekhin:2015byh, SHiP:2018xqw} and DUNE~\cite{Acciarri:2015uup}.  Although the heavy-light neutrino mixing angles $|U_{\ell N}|$ are (or can be) tightly constrained by these experimental data, these limits from $N$ decay products can not be used directly to set limits on the couplings $\lambda_{\alpha\beta}$ in our case, as the $\phi$ decays to invisible neutrinos; therefore, we do not consider these limits here.

\subsection{Invisible $Z$ decays}
\label{sec:Zdecay}

The leptonic scalar $\phi$ could couple to the neutrinos coming from $Z$ decays and thus induce extra contribution to invisible decay width of the $Z$ boson. The analytical calculations of $\Gamma (Z \to \nu_\alpha \nu_\beta \phi)$ are presented in Appendix~\ref{sec:decay1} [cf.~Eq.~\eqref{eqn:Zdecay}]. The combined LEP result for the BR of invisible $Z$ decays has reached the precision of $10^{-4}$~\cite{Acciarri:1998vf, Akers:1994vh, Buskulic:1993ke, Adeva:1991rp}; however, the observed invisible partial width of $Z$ boson $\Gamma_{\rm obs}^{\rm inv} = (499.0 \pm 1.5)$ MeV is below the SM prediction $\Gamma_{\rm SM}^{\rm inv} = (501.44 \pm 0.04)$ MeV at $1.5\sigma$ C.L.~\cite{PDG}.
Then the experimental and theoretical values can be used to set limits on the mass $m_\phi$ and couplings $\lambda_{\alpha\beta}$ via the following formula
\begin{eqnarray}
\left| (\Gamma^{\rm inv}_{\phi} + \Gamma^{\rm inv}_{\rm SM}) - \Gamma^{\rm inv}_{\rm obs} \right| \ < \ P_{\rm CL} \Delta \Gamma^{\rm inv}_{\rm obs} \,,
\end{eqnarray}
with $\Gamma^{\rm inv}_\phi$ being the contribution from the $\phi$-induced decays, and $P_{\rm CL}$ the C.L. parameter. The resultant constraints on the couplings $\lambda_{\alpha\beta}$ at $2\sigma$ C.L. (with $P_{\rm CL} = 2$) are shown in Fig.~\ref{fig:Zinv}. The red and blue shaded regions are respectively for the couplings $|\lambda_{\alpha\alpha}|$ and $|\lambda_{\alpha\beta}|$ (with $\alpha \neq \beta$).

\subsection{Leptonic $W$ decays}
\label{sec:Wdecay}

In an analogous way, the leptonic scalar $\phi$ can also be emitted in the decays of $W \to \ell \nu$ with $\ell = e,\, \mu,\, \tau$. Therefore, the couplings $\lambda_{\alpha\beta}$ can be constrained by leptonic $W$ decay rates using the analytical expression given in Appendix~\ref{sec:decay1} [cf.~Eq.~\eqref{eqn:Wdecay}]. The current LEP uncertainties for the $e$, $\mu$ and $\tau$ flavor leptonic $W$ decays are respectively 0.16\%, 0.15\% and 0.21\% at $1\sigma$ C.L.~\cite{Abbiendi:2007rs, Abdallah:2003zm, Achard:2004zw, Heister:2004wr}. The corresponding limits on $|\lambda_{\alpha\beta}|$ at $2\sigma$ C.L. are shown in Fig.~\ref{fig:Zinv}. The orange and purple lines are respectively for the $e$ and $\mu$ flavors. In addition, we also have the limits from the following $W$-related final states at LEP and LHC, which however turn out to be much weaker and are not shown in Fig.~\ref{fig:Zinv}:
\begin{figure}[t!]
  \centering
  \includegraphics[height=0.4\textwidth]{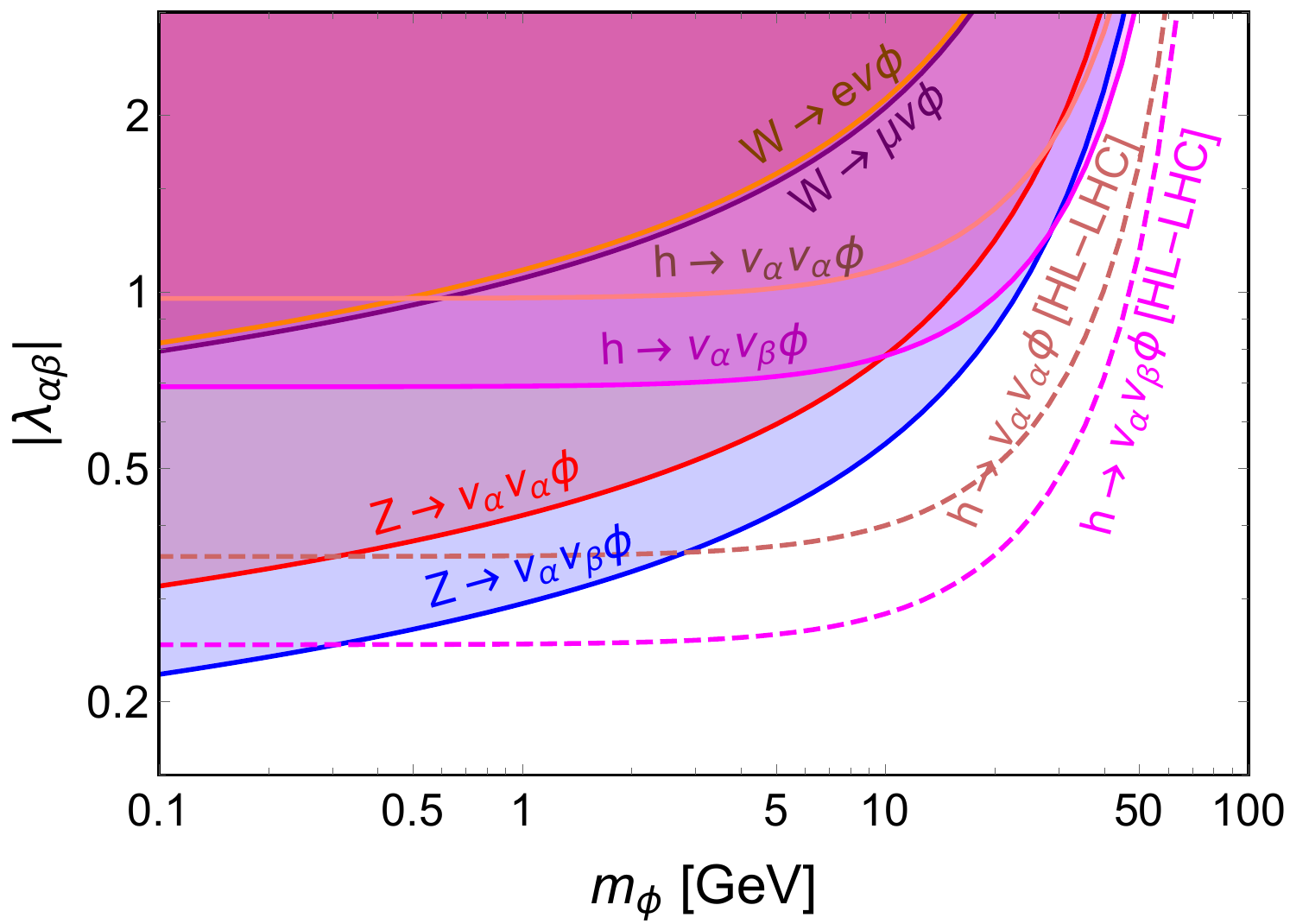}
  \caption{Limits on $|\lambda_{\alpha\beta}|$  (with $\alpha,\,\beta = e,\,\mu$) from invisible $Z$ decay $Z \to \nu_\alpha \nu_\alpha \phi$ (red), $\nu_\alpha \nu_\beta \phi$ with $\alpha\neq\beta$ (blue), the decay $W \to e\nu \phi$ (orange), $\mu \nu \phi$ (purple) and invisible decay of the SM Higgs $h \to \nu_\alpha \nu_\alpha \phi$ (pink), $\nu_\alpha \nu_\beta \phi$ with $\alpha \neq \beta$ (magenta). The data can be found in Table~\ref{tab:limits}, and all the shaded regions are excluded. The dashed pink and magenta lines denote the limits from the prospects of invisible decay of the SM Higgs at the HL-LHC.}
  \label{fig:Zinv}
\end{figure}
\begin{itemize}
  \item The $W$ production cross section times branching fraction for the Drell-Yan process $pp \to W \to \ell \nu$ has been measured at the LHC. The distributions of the transverse momentum $p_T$ of charged lepton, missing energy and the transverse mass of $W$ boson have also been measured by both ATLAS~\cite{Aad:2010yt, ATLAS:2010dda, Aad:2016naf, Aaboud:2016btc, Aaboud:2017svj} and CMS~\cite{Chatrchyan:2011jz, CMS:2011aa, Chatrchyan:2012xt, Chatrchyan:2013mza, Chatrchyan:2014mua, CMS:2015ois}. For sufficiently large couplings $\lambda_{\alpha\beta}$, the leptonic  scalar $\phi$ can be produced from $W \to \ell \nu \phi$ {(or $W^*\to \ell \nu \phi$ for $m_\phi>m_W$)}  and potentially modify the distributions above, which can in principle be used to set limits on scalar mass $m_\phi$ and the couplings $\lambda_{\alpha\beta}$. However, even if we use the current most precise data from Ref.~\cite{Aaboud:2017svj}, the experimental uncertainties are still too large, of order $0.5\%$, when compared to, e.g., those from $Z$ decay (at the level of 0.05\%), and therefore, we cannot obtain stronger  constraints from these distributions.

  \item The charged lepton energy distribution has also been measured in the $W$-pair production process $e^+ e^- \to W^+ W^- \to q\bar{q} \ell \nu$ at LEP~\cite{Abbiendi:2007rs}. In principle, these distributions can be used to set limits on the couplings $\lambda_{\alpha\beta}$. However, the experimental uncertainties again turn out to be too large to put any stronger constraints than those obtained from $W$ and $Z$ decay.

  \item The electron-muon universality has also been tested in the $W$ decay at LHC~\cite{Aaboud:2016btc}. However, the current experimental uncertainties are at $1\%$ level, and therefore, the universality constraints are expected to much weaker than those from the $W$ and $Z$ decay rates.

  \item One can also use other LHC data involving $W$ boson to estimate the constraints on $\lambda_{\alpha\beta}$, such as the top quark pair-production at LHC. However, the inclusive cross section for $pp \to t\bar{t}$ is at least four times smaller than that for single $W$ production at LHC~\cite{top}, and the SM backgrounds for top quark events are more complicated. Therefore we expect the limits from $t \to Wb\to \ell \nu b$ should be significantly weaker than those from $W$ data itself.
\end{itemize}

Comparing the limits on the scalar $\phi$ from the direct production of $\phi$ from $W$ boson decay $pp \to W \to \ell\nu\phi$ at the LHC and the production of $\phi$ via the fusion of same-sign $W$ bosons in Fig.~\ref{fig:diagram} (see Figs.~\ref{fig:LHC} -- \ref{fig:LHC3} for the prospects), one may wonder why the sensitivities from the VBF process is better. The reason is as follows: Although the VBF process has five particles in the final state at the parton level, with the cross section at the level of fb at the 14 TeV LHC (see Fig.~\ref{fig:prod_xsec}), the same-sign di-leptons in the VBF process provide strikingly clean signatures and the corresponding SM backgrounds are not overwhelming. In particular, the kinetic distributions of the leptons, jets and missing energy of the signal are very different from the SM backgrounds, as shown in Figs.~\ref{fig:distributions} and \ref{fig:distributions1}, and the prospects of $\lambda_{\alpha\beta}$ can go up  to 0.95 at LHC and 0.51 at HL-LHC (cf. Table~\ref{tab:results} and Figs.~\ref{fig:LHC}-\ref{fig:LHC3}).

To justify our choice of the VBF process over $pp \to W^{(\ast)} \to \ell\nu\phi$, we divide our parameter space into two regions: (i) $m_{\phi} < m_W$ and (ii) $m_{\phi} \geq m_W$. First, let us discuss $m_{\phi} < m_W$. The on-shell $W$ boson production cross section at the LHC can reach the order of nb, orders of magnitude larger than the VBF process. However, when the kinematic distributions from $W$ decay are used to set limits on the couplings $\lambda_{\alpha\beta}$, we have to compare the impact of the decay $W \to \ell \nu \phi$ and the uncertainties of the $W$ distribution data. As the muon data from $W$ decay is more precise than that for electron and tauon, we take the $p_T$ distributions of muons for the purpose of comparison. The number of muon events in the $i$th $p_T$ bin can be estimated to be
\begin{eqnarray}
\Delta {\cal N}_i \ = \
{\cal L} \times \sigma (pp \to W) \times {\rm BR} (W \to \mu\nu) \times \varepsilon \times C_i \,,
\end{eqnarray}
where ${\cal L}$ is the luminosity, $\sigma (pp \to W)$ is the $W$ production cross-section, ${\rm BR} (W \to \mu\nu)$ is the BR of $W$ decaying into muons, $\varepsilon$ includes the cut and detector efficiency and geometrical acceptance, and $C_i$ is the probability of the muon events lying in the $i$th $p_T$ bin. The presence of the light scalar $\phi$ will lead to the extra contribution to the $p_T$ bins
\begin{eqnarray}
\delta (\Delta {\cal N}_i) \ = \
{\cal L} \times \sigma (pp \to W) \times {\rm BR} (W \to \mu\nu\phi) \times \varepsilon \times C_i^\prime \,,
\end{eqnarray}
with $C_i^\prime$ being the corresponding probability for the $i$th bin. Then the variation in the $i$th muon $p_T$ bin can be estimated to be
\begin{eqnarray}
\frac{\delta (\Delta {\cal N}_i)}{\Delta {\cal N}_i} \ = \
\frac{{\rm BR} (W \to \mu \nu \phi)}{{\rm BR} (W \to \mu \nu)} \times
\frac{C_i^\prime}{C_i} \,.
\end{eqnarray}
In presence of the light scalar $\phi$, the muon $p_T$ distribution will be different, and the $C_i^\prime$'s are expected be different from the $C_i$'s, with $C_i^\prime$'s relatively larger in the bins with relatively small $p_T$.
However, to obtain a quick order of magnitude estimation of the limit on $\lambda_{\mu \mu}$ from the available 7 TeV LHC data in Ref.~\cite{Aaboud:2017svj}, we assume $C_i^\prime \sim C_i$. Then, for the scalar mass $m_\phi = 1$ GeV we find,
\begin{eqnarray}
\label{eqn:variation}
\frac{\delta (\Delta {\cal N}_i)}{\Delta {\cal N}_i} \ \sim \
0.021 \lambda_{\mu\mu}^2 \,.
\end{eqnarray}
The dominant uncertainties for the LHC $W$ data are experimental and physics-modelling systematic uncertainties, which are of order $0.5\%$ for the muon $p_T$ distributions~\cite{Aaboud:2017svj}. Then Eq.~(\ref{eqn:variation}) implies that the LHC $W$ distribution limits on $\lambda_{\alpha\beta}$ can reach  $\sim 0.5$ in the light $\phi$ limit. Although this is at the same order as the VBF prospects but is less constraining than the LEP $Z$ invisible decay-width limits in Fig.~\ref{fig:Zinv} for $m_{\phi} < M_Z$. Hence, we are not showing this limit in Fig.~\ref{fig:Zinv}.

In the estimates above, we do not perform any shape-analysis of either $p_T$ of the lepton, $E_T^{\rm miss}$ or $M_T$ distributions to derive the limit. One may expect that such a shape-analysis will make the bounds coming from $p p \to W \to \ell \nu \phi$ competitive with LEP $Z$ invisible limits, and may even surpass it for $m_{\phi} \gtrsim \mathcal{O}(10$ GeV). We expect to perform a dedicated shape-analysis of this channel in a follow-up study. It is needless to say that if the error bar in the kinematic distributions goes down significantly compared to present $0.5\%$ level at the HL-LHC, this channel might provide comparable or even stronger limits compared to either the invisible $Z$ decay data or the VBF process in this paper.

Now, for $m_{\phi} > m_W$, we perform a simple cut-and-count analysis following the ATLAS study of Ref.~\cite{Aaboud:2016zkn}. We require for an isolated lepton ($\ell = e, \mu$) in the final state with $p_T > 55$ GeV and $|\eta| < 2.5$, $E_T^{\text{miss}} > 55$ GeV, the transverse mass $M_T > 110$ GeV, and $u_T < 30$ GeV, where $u_T$ is the vector sum of transverse momenta of all objects in the event other than the isolated lepton and missing energy. We find that $p p \to W^* \to \ell \nu \phi$ will provide a 95$\%$ C.L. bound on $\lambda_{\alpha \beta} \, (\alpha, \beta = e, \mu)$ in the range of $0.65 - 3.15$ for $m_W \leq m_{\phi} < v$ with the assumption of no systematic errors. However, the signal-to-background ($S/B$) ratio in this channel is $\lesssim 10^{-3}$, making this channel not competitive to our VBF analysis (with $S/B\sim 0.5$, see section~\ref{sec:3.2}) in presence of even small systematic errors.




\subsection{Invisible Higgs decays}\label{sec:Hdecay}
If the effective coupling $\lambda_{\alpha\beta}$ in Eq.~(\ref{eqn:Lagrangian}) originates from the dimension-six operator $(LH)(LH)\phi$, the same operator leads also to the effective couplings of SM Higgs $h$ with the light scalar $\phi$ and neutrinos, i.e.~\cite{Berryman:2018ogk}
\begin{eqnarray}
\label{eqn:Lagrangian2}
{\cal L}_{\rm int} \ \supset \ \frac{\lambda_{\alpha\beta}}{v_{}} h \phi \nu_\alpha \nu_\beta \,,
\end{eqnarray}
which induces the exotic decay of the SM Higgs $h \to \phi \nu \nu$, and the corresponding partial width can be found in Appendix~\ref{sec:decay1} [cf.~Eq.~\eqref{eq:hdecay}]. As the light scalar $\phi$ decays only into neutrinos, such exotic decay of the SM Higgs is completely invisible. The current LHC limits on invisible BR of the SM Higgs is 24\%~\cite{Khachatryan:2016whc}, and the resultant limits on the scalar mass $m_\phi$ and $\lambda_{\alpha\beta}$ are presented in fig.~\ref{fig:Zinv}. The pink and magenta shaded regions are respectively for the cases $h \to \phi \nu_\alpha \nu_\alpha$ and $h \to \phi \nu_\alpha \nu_\beta$ with $\alpha \neq \beta$.
At the HL-LHC, the invisible decay of the SM Higgs can reach a precision of 4.2\%~\cite{Cepeda:2019klc}, and the corresponding limits on the light scalar mass $m_\phi$ and thee couplings $\lambda_{\alpha\beta}$ can be significantly improved, as indicated by the dashed pink and magenta lines in Fig.~\ref{fig:Zinv}.

\subsection{Tauon decay rates}
\label{sec:leptondecay}


If kinematically allowed, the leptonic scalar $\phi$ could also be produced from lepton decays, such as $\mu \to e \nu\nu\phi$ and $\tau \to \ell_\alpha \nu \nu \phi$ (with $\alpha = e,\,\mu$). In the limit of $m_\phi \to 0$ (or $m_\phi \ll m_\mu$) the $\mu$ decay limits are expected to be much stronger than those from $\tau$ decays~\cite{Lessa:2007up}, because the Michel electron spectrum from muon decay has been measured very precisely~\cite{PDG}. However, when the scalar mass $m_\phi \gtrsim 100$ MeV the decay $\mu \to e \nu\nu\phi$ is either kinematically forbidden or highly suppressed, and therefore, we only consider the tauon decays in this subsection. The calculation of partial width for the four-body decays $\Gamma (\tau \to \ell_\alpha \nu \nu \phi)$ is outlined in Appendix~\ref{sec:decay2}.
The partial widths $\Gamma (\tau \to e \nu \nu \phi)$ and $\Gamma (\tau \to \mu \nu \nu \phi)$ are compared to the experimental uncertainties for the leptonic decays ${\rm BR} (\tau \to e \nu \nu)$ and ${\rm BR} (\tau \to \mu \nu \nu)$, which turn out to be $4 \times 10^{-4}$ at the $1\sigma$ level for both $e$ and $\mu$~\cite{PDG}. As the leptonic scalar $\phi$ can be emitted from the $\nu_\alpha$ and/or $\nu_\tau$ fermion lines, all the flavor combinations of $\lambda_{\alpha\beta}$ get constrained by the $\tau$ decay data, including $\lambda_{\tau\tau}$ which is barely constrained by meson decays. It turns out that the tauon decay limits on all flavor combinations $\lambda_{\alpha\beta}$ are roughly the same, as summarized in Figs.~\ref{fig:LHC}--\ref{fig:LHC3}. For a scalar mass of $m_{\phi} = 100$ MeV, it is required that $|\lambda_{\alpha\beta}| \lesssim {\cal O} (1)$. The constraints in the $m_\phi \to 0$ limit are $|\lambda_{\alpha\beta}| \lesssim 0.3$, which is consistent with the numbers given in Ref.~\cite{Lessa:2007up}.

\subsection{Neutrino beam experiments}


As stated in Ref.~\cite{Berryman:2018ogk}, the light leptonic scalar $\phi$ can be emitted from neutrino beams via neutrino-matter scattering such as $\nu_\alpha + p \to \ell_\beta^+ + n + \phi$ (where $p$ and $n$ stand for proton and neutron, respectively). This will affect the charged lepton momentum distributions in the final state, and more importantly,  the charged lepton in this process seems to have the wrong sign due to the emission of lepton-number-charged $\phi$. The magnetized MINOS detector can distinguish the charges of $\mu^\pm$ produced from charge-current neutrino-nucleon interactions~\cite{Rebel:2013vc}, and as a consequence, the coupling $|\lambda_{\mu\mu}|$ is constrained to be smaller than ${\cal O}(1)$ for a 100 MeV scalar mass, as shown by the green shaded region in Fig.~\ref{fig:LHC3}. With more charged-current events collected at DUNE~\cite{Acciarri:2015uup}, the coupling $|\lambda_{\mu\mu}|$ could be probed to a smaller value. For instance, assuming the most aggressive cut of no missing transverse momentum~\cite{Berryman:2018ogk}, the prospects of $|\lambda_{\mu\mu}|$ are expected to be enhanced by one order of magnitude, as shown by the dashed green line in Fig.~\ref{fig:LHC3}. The MiniBooNE limit on $|\lambda_{e\mu}|$ is also considered in Ref.~\cite{Berryman:2018ogk}, which turns to be weaker than those shown in Fig.~\ref{fig:LHC2}.

The NOMAD experiment searched for neutrino oscillation in $\nu_\mu \to \nu_\tau$ channel~\cite{Astier:2001yj}, but no charged-current $\nu_\tau$ event was found, and a limit was imposed on the $\nu_\mu - \nu_\tau$ oscillation probability, which can be translated to a limit on the leptonic scalar mass $m_\phi$ and coupling $|\lambda_{\mu\tau}|$~\cite{Berryman:2018ogk}. The tauon events are difficult to be identified in the DUNE near detector, thus the DUNE prospect of $|\lambda_{\mu\tau}|$ is expected to be weaker than that from NOMAD. Similarly, the limit on the $\nu_e - \nu_\tau$ oscillation probability is weaker than that for $\nu_\mu \to \nu_\tau$.

\subsection{Astrophysical and cosmological limits on neutrino self-interactions}
\label{sec:ic}


A few PeV neutrino events have been observed in the IceCube neutrino  experiment~\cite{Aartsen:2013jdh, Aartsen:2014gkd, Aartsen:2017mau}. These high-energy neutrinos could in principle induce neutrino--neutrino interactions in the early universe, such as the ones mediated by a scalar field $\phi$~\cite{Ng:2014pca, Ioka:2014kca, Barenboim:2019tux} as in our case. For scalar mass $m_\phi \gtrsim 100$ MeV, the $\phi$ mediated neutrino--neutrino interactions are practically effective four-neutrino interactions.  Thus the IceCube PeV neutrino limits on neutrino--neutrino interactions can be translated to a constraint on $|\lambda_{\alpha\beta}|^2/m_\phi^2$ as shown by the shaded blue region in Figs.~\ref{fig:LHC}--\ref{fig:LHC3} which are universal for all the three neutrino flavors. Future IceCube data will improve the limits significantly~\cite{Ng:2014pca, Ioka:2014kca}, as shown by the dashed blue line in Figs.~\ref{fig:LHC}--\ref{fig:LHC3}.\footnote{Here we do not include the resonance effect for the IceCube limits, which depend on the neutrino masses and the neutrino energies in IceCube data. When the resonance effect is taken into account, the current IceCube limits on $\lambda_{\alpha\beta}$ and the future prospects will be improved by up to  two orders of magnitude for respectively the scalar mass ranges of $m_\phi \sim$ MeV -- 10 MeV and $\sim$ MeV -- GeV (see Fig.~1 of Ref.~\cite{Ioka:2014kca}). However, our main results and conclusions of the LHC prospects in this paper will not be affected by including the resonance effect.}

The $\phi$-mediated self-interactions of neutrinos also have some effects in the early Universe. In particular, neutrino free streaming will alter the CMB temperature power spectrum~\cite{Hannestad:2004qu, Bell:2005dr, Cyr-Racine:2013jua}. Current precision cosmological data have excluded the effective coupling $G_{\rm eff} \simeq |\lambda_{\alpha\beta}|^2/m_\phi^2 \gtrsim 2.5 \times 10^7 G_F$~\cite{Basboll:2008fx, Cyr-Racine:2013jua, Archidiacono:2013dua, Lancaster:2017ksf, Oldengott:2017fhy, Kreisch:2019yzn}. This constraint is however weaker than the IceCube constraints discussed above, and hence, not shown in Figs.~\ref{fig:LHC}--\ref{fig:LHC3}.

\subsection{Other limits}
\label{sec:others}

As the leptonic scalar $\phi$ decays invisibly into light neutrinos, it can be constrained by the searches of light DM $\chi$ in the high-intensity beam-dump experiments. For instance, the dark photon $A'$ has been searched for in the NA64 experiment~\cite{NA64:2019imj}  via the electron-nuclei scattering process, with $A'$ subsequently decaying  into a pair of DM particles:
\begin{eqnarray}
\label{eqn:NA64}
e {\cal N} \ \to \ e {\cal N} A', \;\;\; A' \ \to \ \chi\chi \,,
\end{eqnarray}
with ${\cal N}$ being the incident nuclei. The presence of $\phi$ in our case would induce the process
\begin{eqnarray}
\label{eqn:NA64:2}
e {\cal N} \ \to \ e {\cal N} \nu \nu + \phi \,,
\end{eqnarray}
via the fusion of two $Z$ bosons, similar to that shown in Fig.~\ref{fig:diagram} (with the $W$ bosons and charged leptons replaced respectively by $Z$ and neutrinos). The final states are the same as that for the DM searches, i.e. the scattered nuclei and electron plus significant missing energy. Therefore the limits on the dark photon mass $m_{A'}$ and the kinetic mixing $\varepsilon$ of dark photon with the SM photon from the NA64 experiment can be recast into limits on the mass $m_\phi$ and $\lambda_{\alpha\beta}$ coupling in our case. However, the center-of-mass energy $E_{\rm cm}$ of the beam-dump experiment is expected to be much lower than the $Z$ boson mass $m_Z$, thus the $\phi$-induced process cross-section is highly suppressed by the ratio $(E_{\rm cm}/m_Z)^8$. Furthermore, comparing the three-body and five-body phase spaces in the two processes in Eq.~(\ref{eqn:NA64}) and (\ref{eqn:NA64:2}), the $\phi$ production process is further suppressed by a factor of ($4\pi)^4$. As a result, the NA64 experiment can not provide any limit on the $\phi$ scalar. Although the LDMX experiment has a larger intensity than NA64, but the center-of-mass energy is lower~\cite{Akesson:2018vlm}, and we can not get any sensitivity of $m_\phi$ and the couplings $\lambda_{\alpha\beta}$.

The process $e {\cal N} \to e {\cal N} A'$ has also been searched for in the DarkLight experiment for a lighter dark photon, but he dark photon mass $m_{A'} < 100$ MeV~\cite{Balewski:2014pxa}. The searches of dark photon in the process $e^+ e^- \to \gamma + A'$ with $A' \to {\rm inv.}$ has been performed in the BaBar experiment~\cite{Lees:2017lec}, and also proposed in VEPP-3~\cite{Wojtsekhowski:2012zq, Wojtsekhowski:2017ijn}. The final state in this case is a mono-energetic photon plus missing energy. The $\phi$ can not induce such signals in our case, so these limits can not be used on the $\lambda_{\alpha\beta}$ couplings.

When the leptonic scalar $\phi$ is light, with $m_\phi \lesssim 100$ MeV, its couplings to neutrinos could also induce very rich phenomena, some of which lead to very stringent limits on $\lambda_{\alpha\beta}$. For such light scalars the prospects of the couplings $\lambda_{\alpha\beta}$ at LHC and HL-LHC are almost independent of $m_\phi$, as can be seen in Figs.~\ref{fig:LHC}--\ref{fig:LHC3} (see section~\ref{sec:prospects}) and cannot compete with the low energy processes. Therefore, we restrict the $\phi$ mass to 100 MeV and do not show these low-energy limits in Figs.~\ref{fig:meson}--\ref{fig:Zinv} and \ref{fig:LHC}--\ref{fig:LHC3}. Nevertheless, for completeness, we list some of these processes below:
\begin{itemize}
  \item {\it Muon decay}: As discussed in section~\ref{sec:leptondecay}, $\phi$ can be emitted from tree-level decay $\mu \to e \nu \nu \phi$. As a result of the precise $\mu$ decay data, for sufficiently light $\phi$, the limits from $\mu$ decay are expected to be much more stringent than those from $\tau$ decays. In addition, the electron~\cite{Derenzo:1969za, Bayes:2011zza} and neutrino~\cite{Armbruster:1998qi, Armbruster:2003pq} spectra could be altered in presence of $\phi$, which can also be used to set limits on $\lambda_{\alpha\beta}$.

  \item {\it Tritium decay}: If the scalar mass $m_\phi \lesssim {\cal O} (10 \, {\rm eV})$, it can be produced from tritium decay in the process $^3 {\rm H} \to \, ^3 {\rm He}^+ + e^- + \nu + \phi$~\cite{Arcadi:2018xdd}, and this process can be probed in the KATRIN experiment~\cite{Osipowicz:2001sq, Angrik:2005ep}.

  \item {\it $0\nu\beta\beta$ decay}: The coupling of $\phi$ to electron neutrinos contributes to $0\nu\beta\beta$ decays via the process $(Z,A) \to (Z+2,A)e^- e^- \phi$ if the mass $m_\phi \lesssim {\cal O} ({\rm MeV})$ -- the typical $Q$-value for the relevant nuclei. This is strongly constrained by the searches of Majoron emission in $0\nu\beta\beta$ decay experiments like NEMO-3 using $^{100}$Mo~\cite{Arnold:2013dha, Arnold:2015wpy, NEMO-3:2019gwo} and $^{150}$Nd~\cite{Arnold:2016qyg} nuclei, as well as  KamLAND-Zen~\cite{Gando:2012pj} and EXO-200~\cite{Albert:2014fya} using $^{136}$Xe. Somewhat weaker limits were also obtained by NEMO-3 using $^{48}$Ca~\cite{Arnold:2016ezh} and $^{82}$Se~\cite{Arnold:2018tmo}, as well as by GERDA using $^{76}$Ge~\cite{Agostini:2015nwa}.

  \item {\it Supernovae}: A light $\phi$ can be produced abundantly in the supernova core if its mass $m_\phi \lesssim {\cal O} (30 \, {\rm MeV})$ -- the typical core temperature of supernovae. The couplings $|\lambda_{\alpha\beta}|$ can be constrained from both the luminosity and deleptonization arguments~\cite{Choi:1987sd, Farzan:2002wx, Heurtier:2016otg}.

  \item {\it CMB and BBN}: As a light particle, $\phi$ itself contributes to the relativistic degrees of freedom $N_{\rm eff}$ if the mass $m_\phi \lesssim 100$ keV~\cite{Huang:2017egl}. The current precision cosmological data $\Delta N_{\rm eff} = 0.18$ at $1\sigma$ C.L.~\cite{Aghanim:2018eyx} has excluded a large parameter space for such light leptonic scalar mass $m_\phi$ and the couplings $|\lambda_{\alpha\beta}|$. Similarly, the big-bang-nucleosynthesis (BBN) constraints rule out $m_\phi\lesssim 0.2$ MeV for sizable couplings $\lambda_{\alpha\beta}$, as long as they allow $\phi$ particles to thermalize at BBN temperature~\cite{Ahlgren:2013wba,Escudero:2019gfk}.


  \item {\it Neutrino decay}: For sufficiently light $\phi$, the heavier neutrinos might decay via $\nu_j \to \nu_i + \phi$ with the mass indices $i,\,j = 1,\, 2,\, 3$ and $i < j$. Therefore we can impose stringent bounds on the leptonic scalar mass $m_\phi$ and the $\lambda_{ij}$ couplings from the solar neutrino data~\cite{Beacom:2002cb, Berryman:2014qha, Picoreti:2015ika, Aharmim:2018fme, Funcke:2019grs}. There are also constraints from atmospheric and long baseline experiments~\cite{GonzalezGarcia:2008ru, Gomes:2014yua, Choubey:2018cfz}. The CMB limits on neutrino free streaming could also set limits on neutrino decays, as long as the mediator is lighter than neutrino mass and the non-diagonal couplings $\lambda_{ij}$ are non-vanishing~\cite{Hannestad:2005ex, Basboll:2008fx, Archidiacono:2013dua}.
\end{itemize}

\section{Prospects at the LHC and HL-LHC}
\label{sec:LHC}

\begin{figure}
     \centering
     \includegraphics[height=0.4\textwidth]{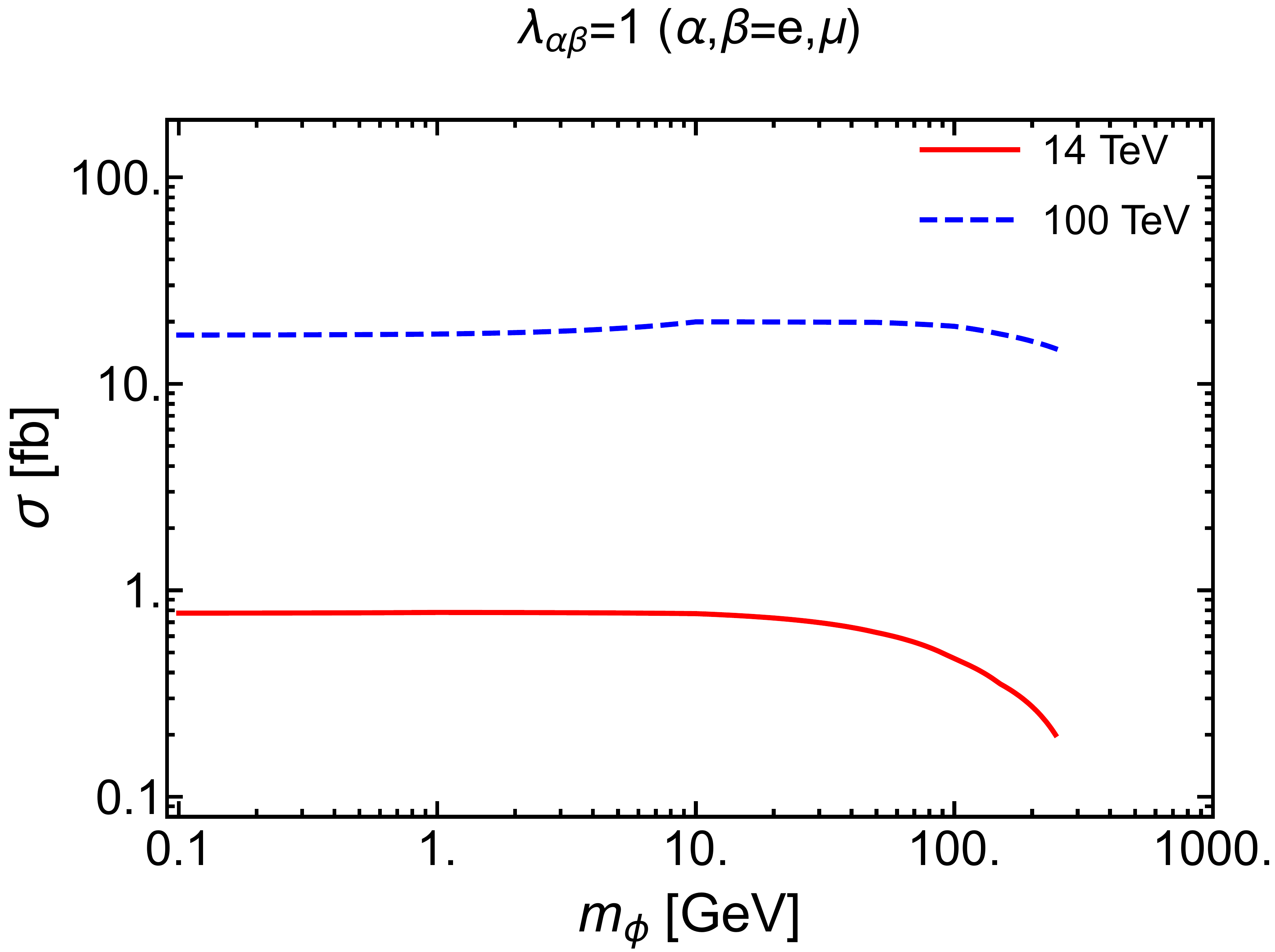}
     \caption{Production cross section of the $ p p \to \ell_\alpha^\pm \ell_\beta^\pm\ \phi\ jj$ process at  $\sqrt{s} = 14$ TeV and 100 TeV, as a function of the mass of $\phi$, with the Yukawa couplings $\lambda_{\alpha\beta} = 1$ ($\alpha,\, \beta = e,\,\mu$). For different coupling values, the corresponding cross sections can be obtained from the scaling $\sigma\propto |\lambda_{\alpha\beta}|^2$. We stop at $m_\phi=v$ beyond which the EFT approach used to define the effective $\nu\nu\phi$ coupling in Eq.~\eqref{eqn:Lagrangian} may not be reliable.}
     \label{fig:prod_xsec}
 \end{figure}

At high-energy colliders, $W$, $Z$ and $h$ decays can give rise to the leptonic scalar $\phi$ via its couplings to neutrinos if kinematically allowed ($m_\phi < M_{W,Z}$), as discussed in sections~\ref{sec:Zdecay} to \ref{sec:Hdecay}. Instead, in this section, we explore the direct production of $\phi$ at the LHC that could potentially extend the reach to higher masses.  At the leading order, $\phi$ can be produced in the VBF  processes $W^\pm W^\pm \to \ell^\pm \ell^\pm \phi,$ leading to the characteristic signal of same-sign dileptons at hadron colliders:
\begin{align}
     p p \ \to \ \ell_\alpha^\pm \ell_\beta^\pm\ \phi\ jj \, ,
     \label{eq:vbf}
\end{align}
where $\alpha, \beta = e, \mu$ are the flavor indices. In a VBF process, two incoming quarks can emit virtual same-sign $W$ bosons, which then interact to produce a pair of same-sign leptons via $t/u$-channel neutrino exchange. The leptonic scalar $\phi$ is irradiated by the $t/u$-channel neutrino. A representative diagram of the process is shown in Fig.~\ref{fig:diagram}.
The choice of our VBF cuts on tagging jets ensures that the same-sign $W$-pair system is boosted, and hence highly energetic. We find that the same-sign leptons carry most of this energy. Consequently, the production cross section of the $\ell^\pm \ell^\pm\ \phi\ jj$ process is not sensitive to a large range of $m_\phi$. In Fig.~\ref{fig:prod_xsec}, we show the variation of the production cross section of the above process~\eqref{eq:vbf} at the $\sqrt{s} = 14$ TeV LHC as a function of $m_{\phi}$ in solid red. In a broad range of mass, the cross section is of ${\cal O}$(1 fb).
 It is evident from Fig.~\ref{fig:prod_xsec} that the creation of $\phi$ at the LHC via VBF processes starts feeling the effect of $\phi$ mass only for $m_{\phi} \gtrsim 10$ GeV. For comparison, we also show the cross section curve of the process for a 100 TeV $pp$ collider in dashed blue. The production rate will be increased by about a factor of 20.

 We only consider $\ell = e, \mu$ in the present study for simplicity. We will comment on the impact of including signals from leptonic $\tau$ decays for our results. However, including hadronic $\tau$ decays in the analysis will require careful examination of a different set of SM backgrounds dominated by $\tau_h$ charge misreconstruction processes, which we postpone for a future study.

\subsection{SM backgrounds and simulation details}

Our strategy to search for $\phi$ is based on two steps. First, we use the distinct features of VBF processes to reduce non-VBF QCD  backgrounds. A VBF process is characterized by two back-to-back energetic jets in the forward/backward region of the detector, with large di-jet invariant mass, and significant separation in rapidity $|\Delta y_{j_1j_2}|$. To select the VBF topology we roughly follow the strategy used in a recent ATLAS $W^{\pm}W^{\pm}jj$ analysis~\cite{Aaboud:2019nmv}. Finally, we impose stringent cuts on the transverse momentum of the leptons, and the azimuthal separation between the leading lepton and transverse missing energy ($E_T^{\text{miss}}$) to suppress the irreducible EW $W^{\pm} W^{\pm} jj$ background.


The dominant SM background processes for our chosen final state are
\begin{itemize}
  \item the EW process $pp \to W^\pm W^\pm jj \to jj \ell_\alpha^\pm \ell_\beta^\pm \nu \nu$,
  \item the QCD process $pp \to W^\pm W^\pm jj \to jj \ell_\alpha^\pm \ell_\beta^\pm \nu \nu$,
  \item $pp \to W^\pm Z jj \to jj \ell_\alpha^\pm \ell_\beta^\pm \ell_\beta^\mp \nu$,
\end{itemize}
with the lepton flavor indices $\alpha,\, \beta = e,\, \mu, \tau$. One should note that although we do not consider light leptons coming from $\tau$ decays for the signal, we do include them for backgrounds. The $W^\pm Z jj$ background is generated inclusively and consists of both QCD and EW processes.
Both the EW and QCD $W^\pm W^\pm jj$ processes have the same final state as the $\phi$-induced signal, i.e., a pair of same-sign dilepton, two hard jets and large $E_T^{\text{miss}}$. At the leading order, the EW $W^\pm W^\pm jj$ background is dominated by the vector-boson scattering  $W^\pm W^\pm \to W^\pm W^\pm$, mediated by a $t$-channel $Z/\gamma$, which 
has recently been observed by both ATLAS~\cite{Aaboud:2019nmv} and CMS~\cite{  Sirunyan:2017ret}. On the other hand, the QCD $W^\pm W^\pm jj$ background is mediated by a $t$-channel gluon. As we will see soon, the QCD $W^\pm W^\pm jj$ background is effectively suppressed by the VBF cuts. For the $W^\pm Z jj$ background, one of the charged leptons coming from the $Z$ decay is missed by the detectors, and we are left with only two isolated leptons in the final state.

There are also some sub-leading backgrounds, such as the charged leptons from heavy-flavor hadron decays, jets misidentified as leptons, backgrounds coming from lepton charge misidentification and the $V\gamma$ production with photon misidentified as electron~\cite{Aaboud:2019nmv,   Sirunyan:2017ret}. All these fake-lepton backgrounds due to detector effects are difficult to simulate reliably within our simulation setting, although a theorist's version of ``fake tagging'' can in principle be performed using the public ATLAS/CMS detector performance studies listing fake rates, in the same manner as heavy flavor jet mistagging, with event re-weighting, or with a global scaling factor; see e.g.~Refs.~\cite{Alva:2014gxa, Alvarez:2016nrz}. For simplicity, we will not consider these fake lepton backgrounds in our analysis. In addition, we neglect the impact of normalization and shape changes at NLO+parton shower in multiboson backgrounds and of continuum contributions, underlying events, multiple parton interaction and pileup effects. Our estimation is that they can contribute up to $20\%$ systematic uncertainty after the VBF cuts, as suggested in  Refs.~\cite{Aaboud:2019nmv, Sirunyan:2017ret} (see also Refs.~\cite{Ballestrero:2018anz, Jager:2018cyo, Jager:2020hkz}). In particular, we expect that the hard lepton $p_T$ and $|\Delta \phi_{\ell_1, E_T^{\text{miss}}}|$ cuts will suppress them significantly due to their efficiency in removing the dominant SM backgrounds (see Table~\ref{tab:cutflow}). However, a detailed experimental study is warranted to properly analyze the relevance of these neglected backgrounds for our proposed signal. In addition, other non-prompt backgrounds like $ZZ, VVV$ and $t \bar{t} V \, (V=W,Z)$ contribute $< 2 \%$ to the total background after the VBF cuts~\cite{Aaboud:2019nmv}, and are not considered here.

The only BSM inputs for our signal estimation at the LHC are the coupling $\lambda_{\alpha\beta}$ and mass $m_\phi$. We add the effective Lagrangian of Eq.~(\ref{eqn:Lagrangian}) to the SM Lagrangian in  {\tt FeynRules}~\cite{Alloul:2013bka} to generate the UFO file, which is used in the simulation of signal events.
We simulate the signal and background events by using {\tt MadGraph5$\_$v2$\_$5$\_$4}~\cite{Alwall:2014hca} with the {\tt NNPDF2.3} LO parton distribution functions~\cite{Ball:2014uwa}. We pass the simulated events to {\tt Pythia8}~\cite{Sjostrand:2007gs} for showering and hadronization, and subsequently to {\tt Delphes-3.4.1}~\cite{deFavereau:2013fsa} for detector simulation. The $W^{\pm}, Z$ bosons in SM backgrounds and the scalar $\phi$ in the signal are decayed to leptons and neutrinos by using the {\tt Madspin}~\cite{Artoisenet:2012st} module of {\tt MadGraph5}.

In our detector simulation with {\tt Delphes}, electrons and muons are identified with $p_T > 10$ GeV and $|\eta| \leq 2.5$. While the muon efficiency is $95\%$ for the entire range of $|\eta|$, the electron efficiency is $95\% \, (85\%)$ for $|\eta| \leq 1.5 \,\, (1.5 < |\eta| \leq 2.5 )$. The lepton isolation is parameterized by $I_{\rm rel} < 0.12 \, (0.25)$ for electron (muon), where $I_{\rm rel}$ is the ratio of the $p_T$ sum of objects (tracks, calorimeter towers, etc) within a
$\Delta R = \sqrt{(\Delta \eta)^2 + (\Delta \phi)^2} = 0.4$ cone around a candidate, and the candidate's $p_T$. Jets are clustered using the anti-$k_T$ algorithm~\cite{Cacciari:2008gp} with cone radius 0.5 and $p_T > 20$ GeV. We use the default $b$-tagging algorithm of {\tt Delphes} where the {$b$-tagging efficiency is just above $70 \%$ for transverse momenta between 85 and 250 GeV, with a mistag rate $\lesssim 2 \% \, (20 \%)$, coming from light jets, i.e. from gluon and up, down, strange (and charm) quarks, over the same $p_T$ range}.

We perform a calibration study to check the lepton isolation performance of {\tt Delphes} against the $\ell^{\pm} \ell^{\pm} + 2 j + E_T^{\text{miss}}$ final state analysis carried out by the ATLAS collaboration in Ref.~\cite{Aaboud:2019nmv}. For the EW $W^{\pm} W^{\pm} jj$ process we obtain 64 events with 36.1 fb$^{-1}$ of integrated luminosity at $\sqrt{s}= 13$ TeV. The corresponding signal yield prediction by the ATLAS collaboration is $60 \pm 11$. For different light lepton flavor combination channels also, we agree with the ATLAS prediction within the error bars.

We finish the discussion on our simulation set-up with one final comment on SM background generation. For fast computation, we generate the $WZ$+jets background for our analysis with exactly two partons, and include both (mixed) QCD and (pure) EW contributions, which at leading order are respectively of order $\mathcal{O}(\alpha^2_s \alpha^2_{\rm EW})$ and $\mathcal{O}( \alpha^4_{\rm EW})$ in the cross section. We do not generate the above background inclusively and do not perform any jet-matching. However, we did check that after VBF cuts the $WZjj$ cross section is within $10\%$ of the inclusive $WZ$+jets cross section. In contrast, both the EW and QCD same-sign $W$ pair is produced at the LHC in association with two partons at LO and do not require inclusive generation.

\subsection{Selection cuts and cross sections} \label{sec:3.2}

\begin{figure}[t!]
  \centering
  \includegraphics[height=0.33\textwidth]{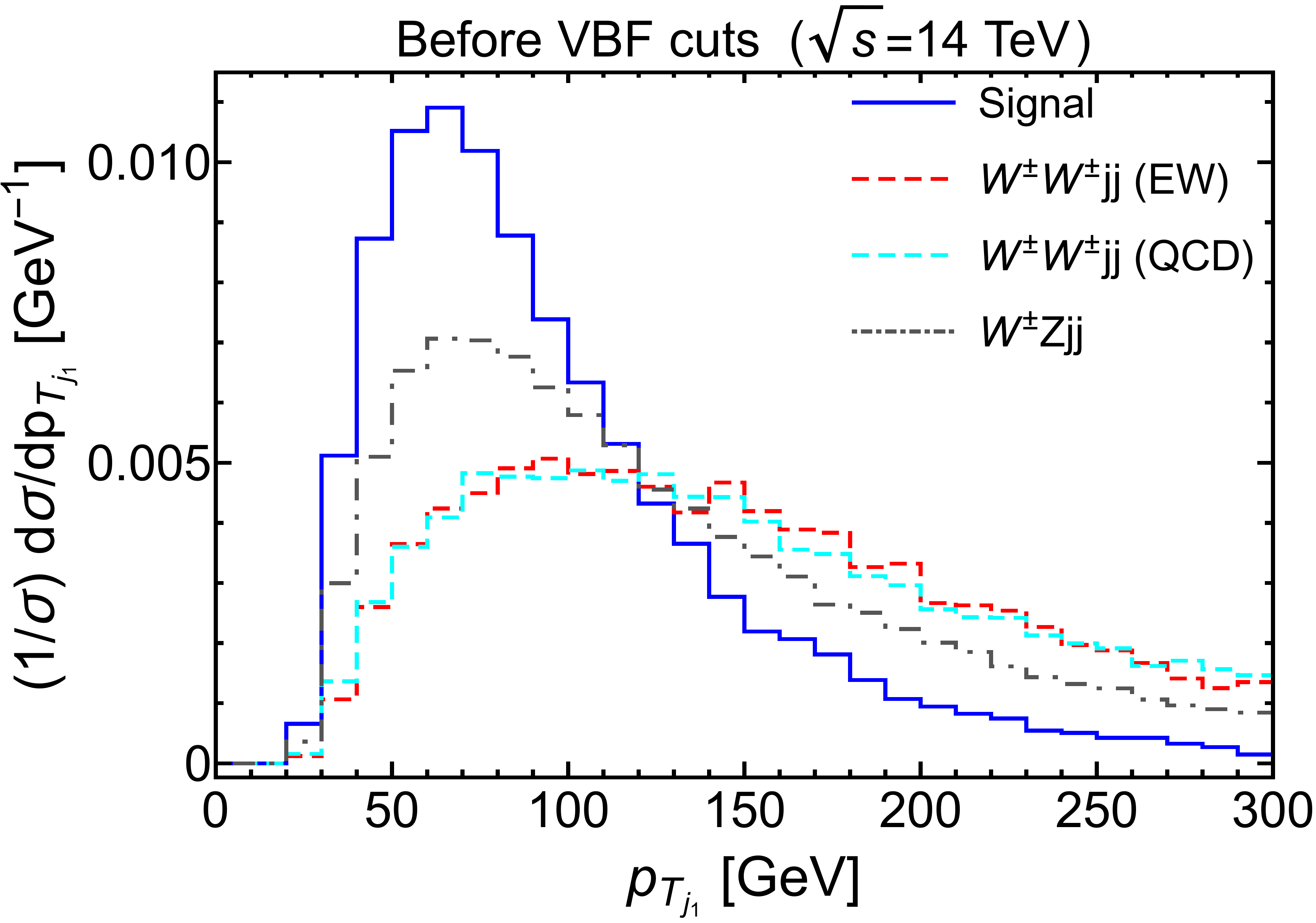}
  \includegraphics[height=0.33\textwidth]{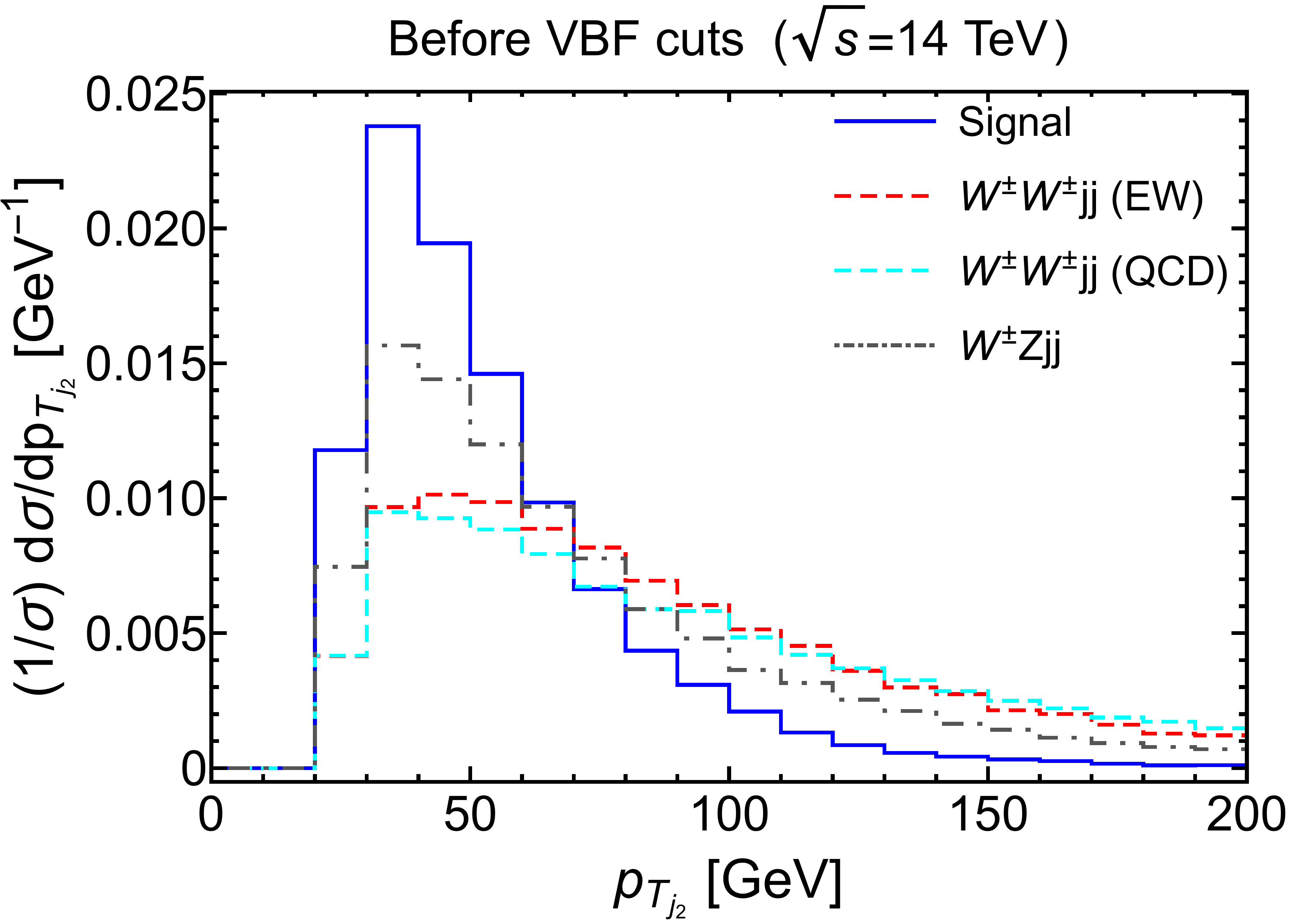}
  \includegraphics[height=0.33\textwidth]{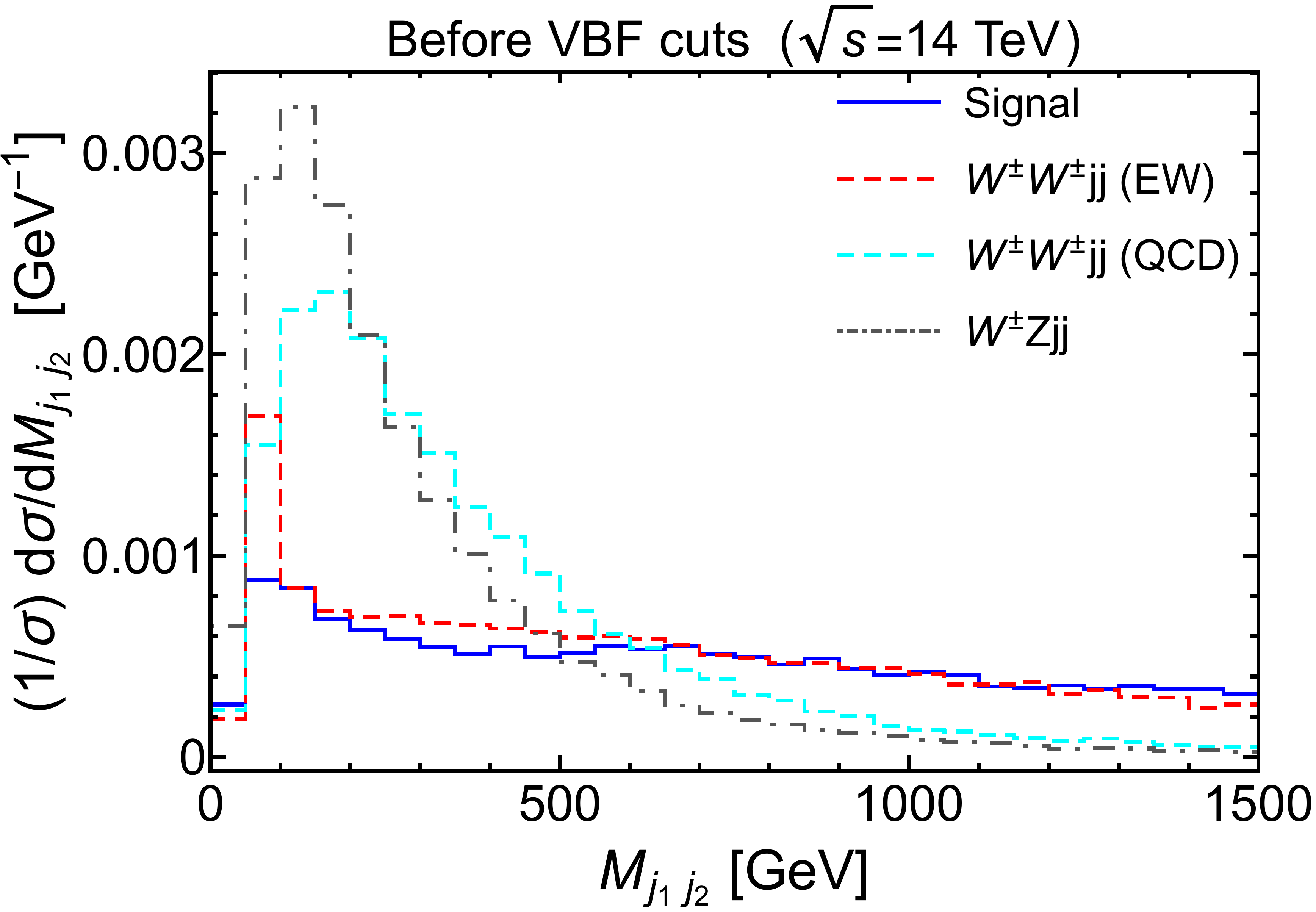}
  \includegraphics[height=0.33\textwidth]{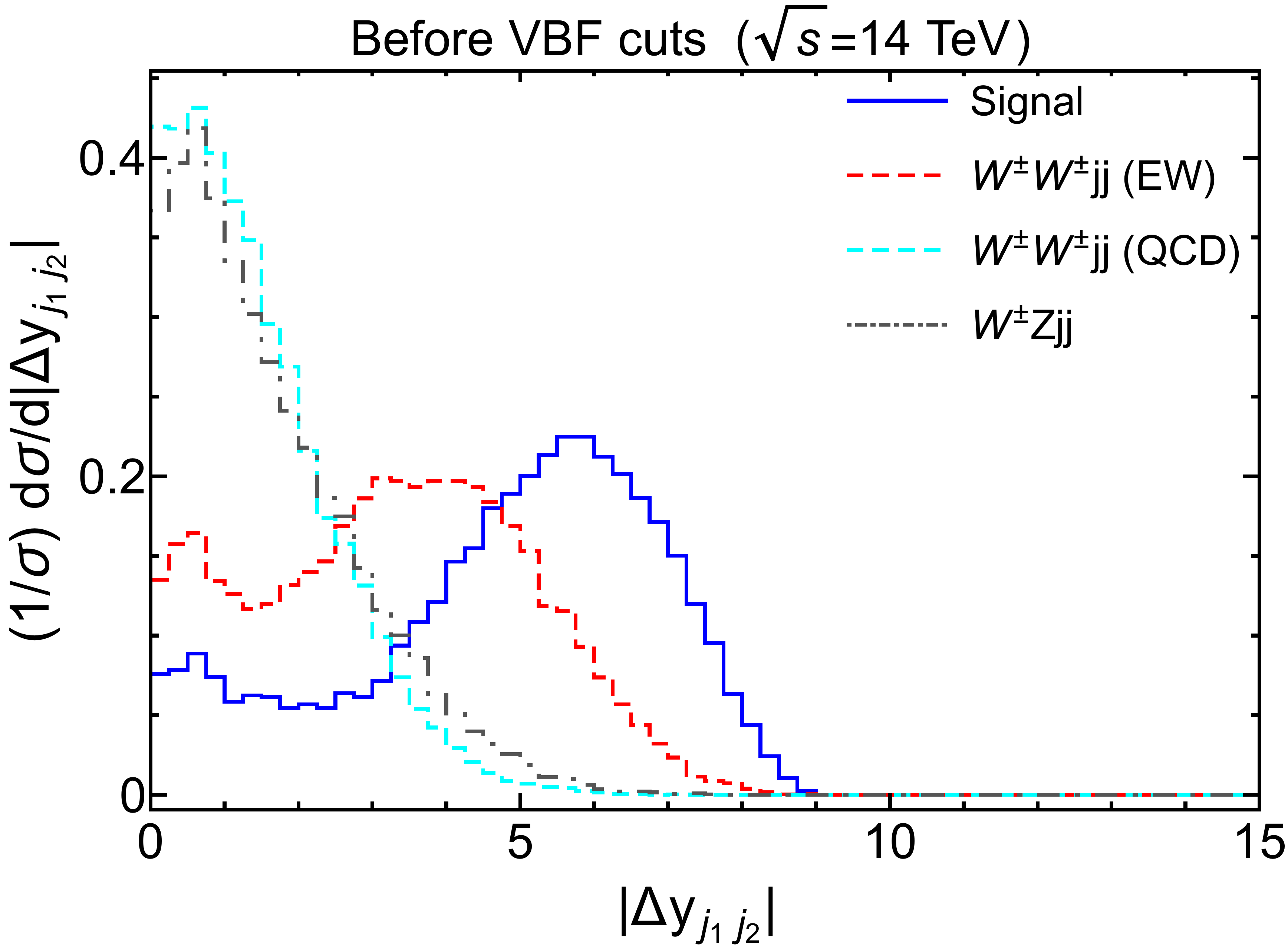}
  \caption{Jet kinematic distributions of the signal and SM backgrounds $W^\pm W^\pm jj$ (EW), $W^\pm W^\pm jj$ (QCD) and $W^\pm Z jj$ (QCD+EW) before VBF cuts. The top left, and right panels are respectively for the $p_T$ distributions of the leading jet $j_1$ and the sub-leading jet $j_2$, and the lower left and right panels are respectively for the invariant mass $M_{j_1j_2}$ and the rapidity separation $|\Delta y_{j_1j_2}|$ of the two jets.}
  \label{fig:distributions}
\end{figure}

\begin{figure}
  \centering
  \includegraphics[height=0.33\textwidth]{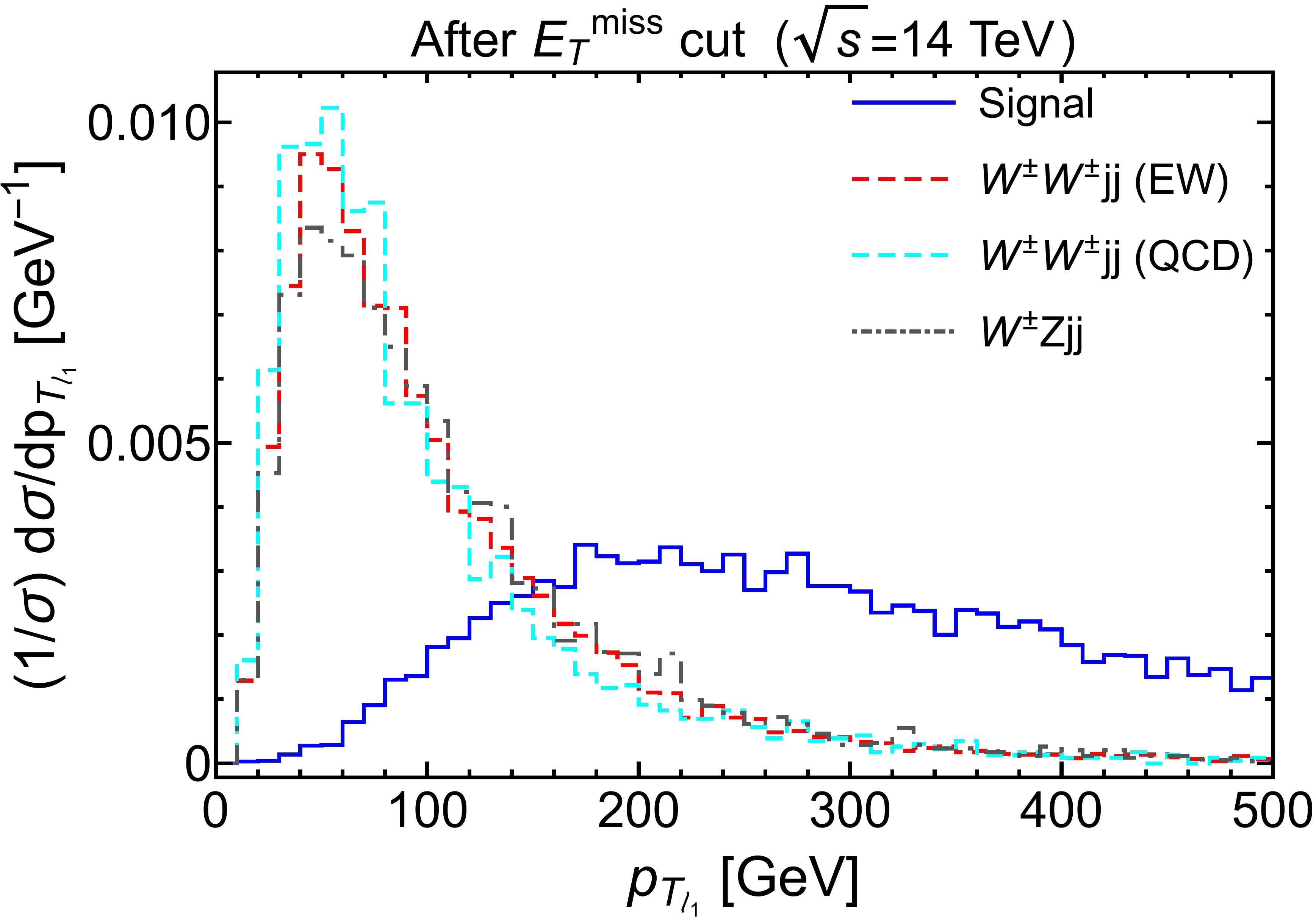}
  \includegraphics[height=0.33\textwidth]{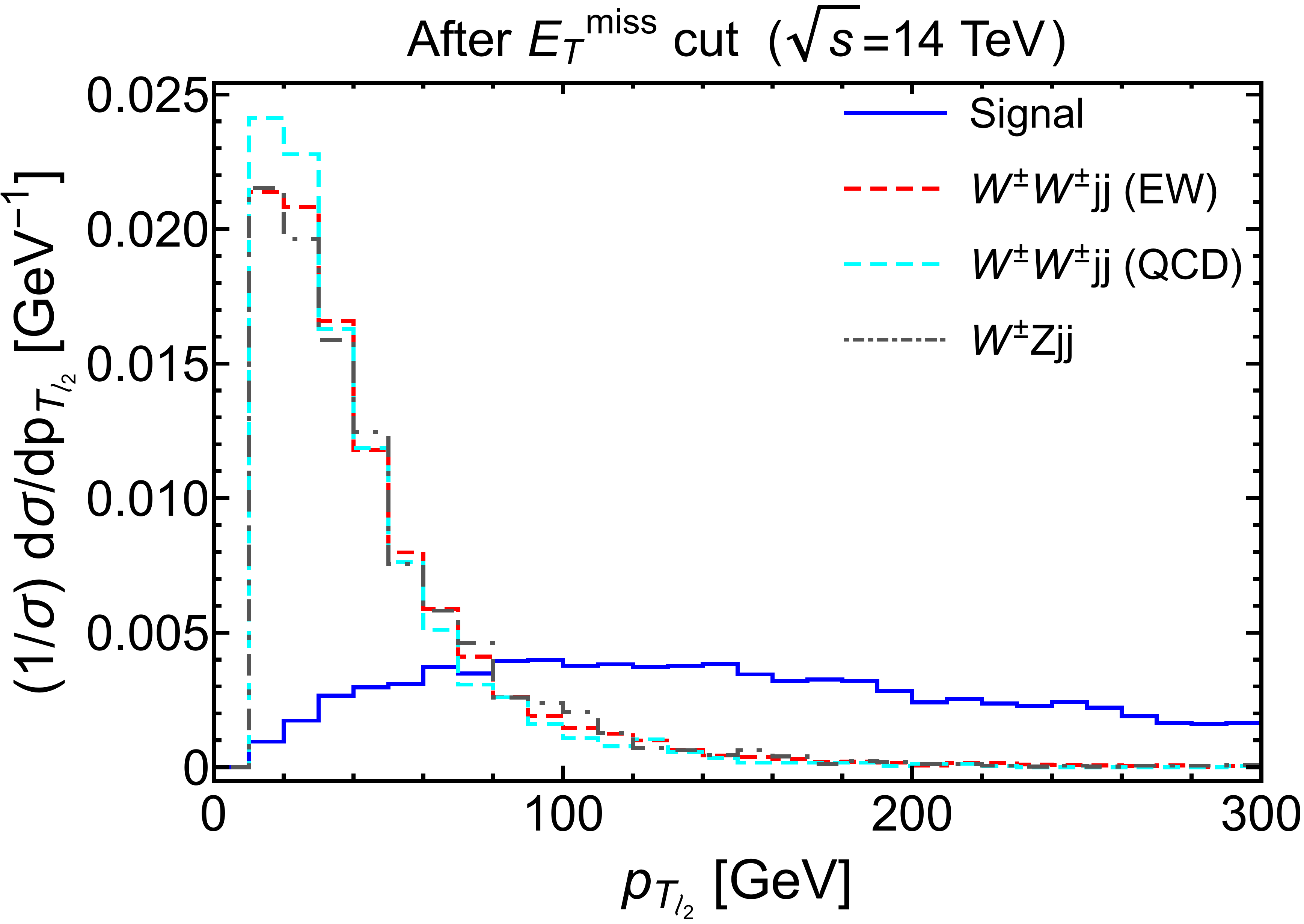}
  \includegraphics[height=0.33\textwidth]{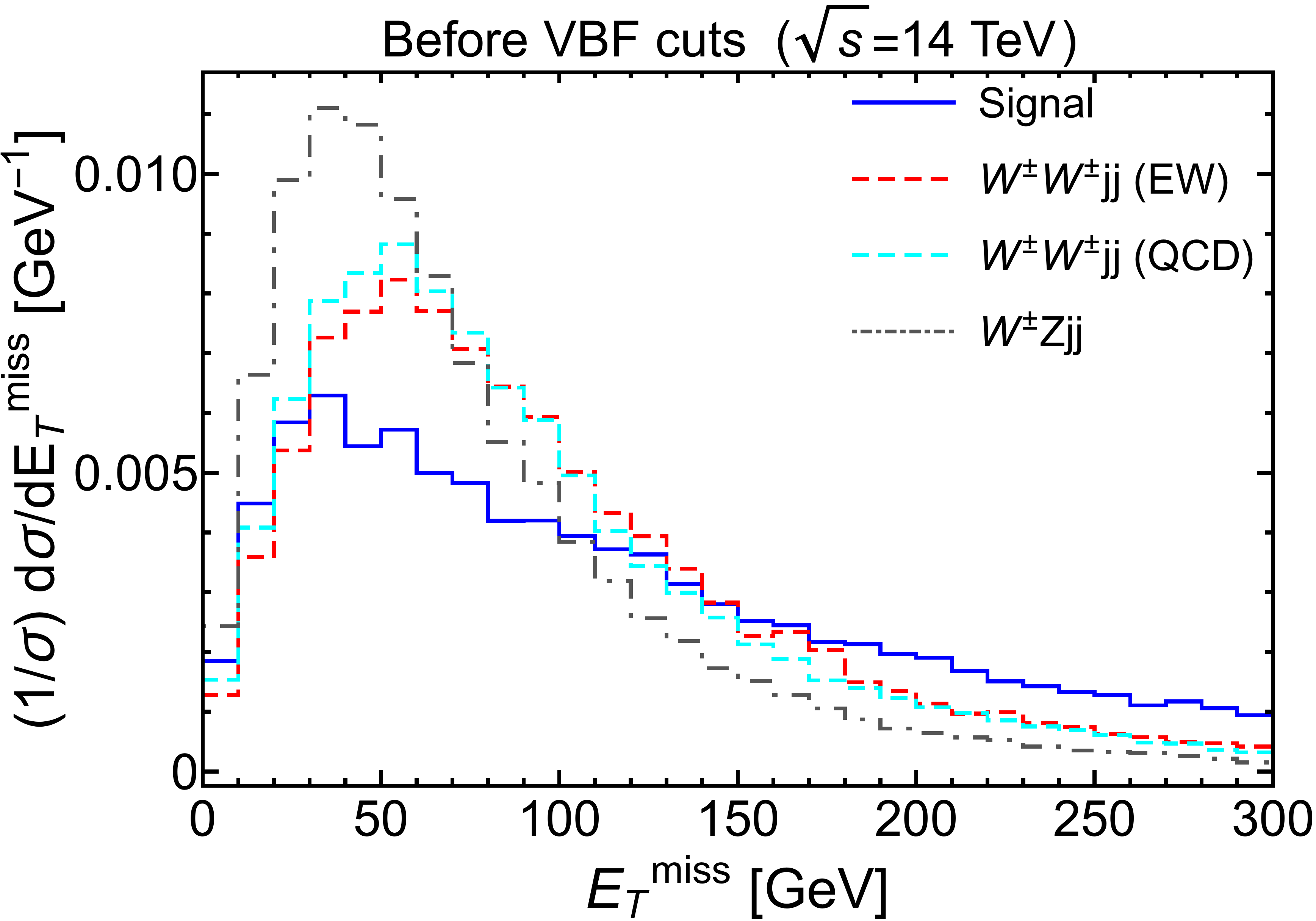}
  \includegraphics[height=0.33\textwidth]{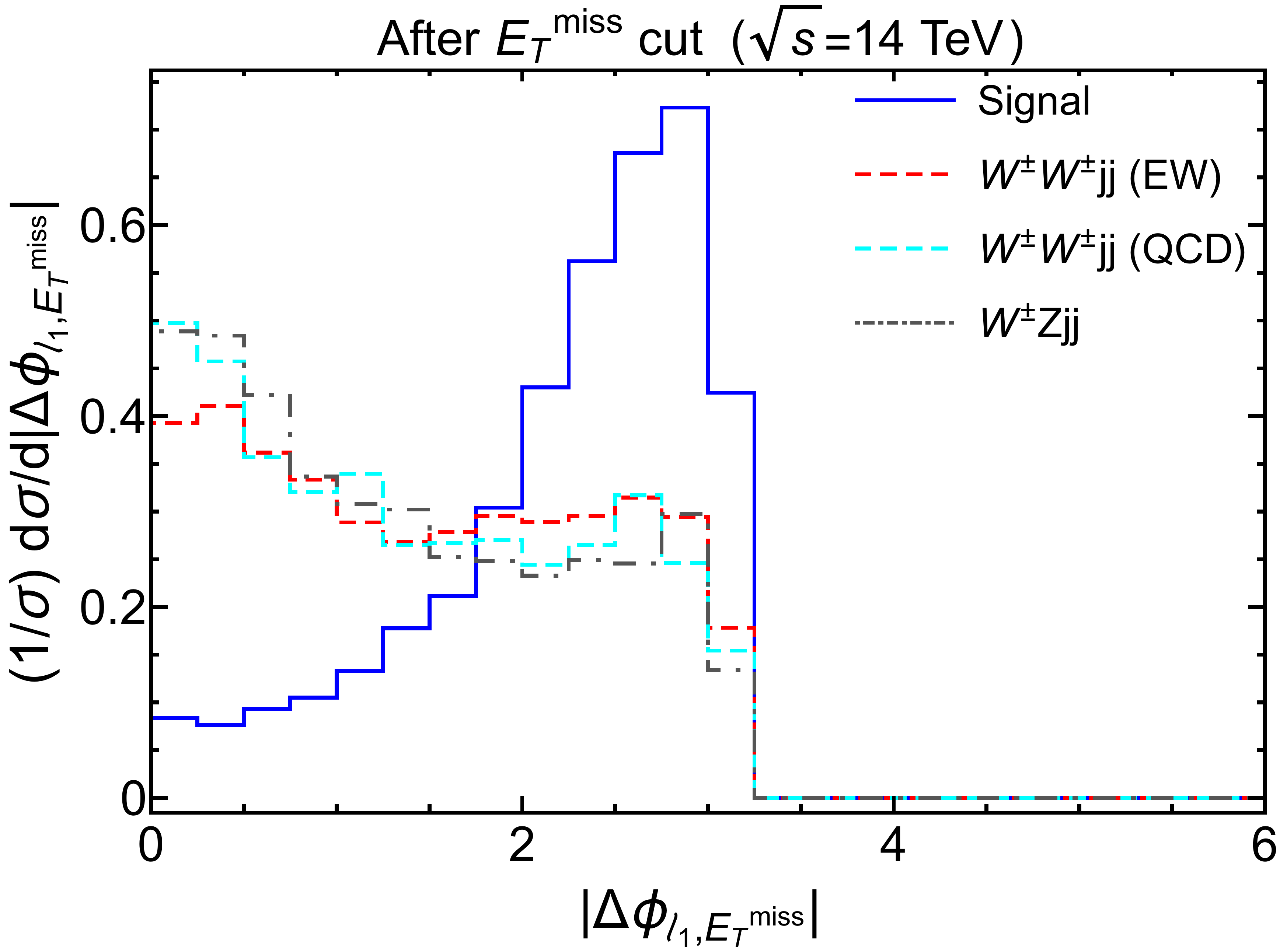}
  \caption{Kinematic distributions of the signal and SM backgrounds $W^\pm W^\pm jj$ (EW), $W^\pm W^\pm jj$ (QCD) and $W^\pm Z jj$ after the  $E_T^{\text{miss}}$ cut. The top left and right panels are respectively for the $p_T$ distributions of the leading lepton $\ell_1$ and the sub-leading lepton $\ell_2$, and the lower left and right panels respectively for the missing transverse energy $E_T^{\rm miss}$ and the angular separation $|\Delta \phi_{\ell_1 E_T^{\rm miss}}|$. Only the $E_T^{\text{miss}}$ distribution is shown before VBF cuts.}
  \label{fig:distributions1}
\end{figure}

Next, using the reconstructed leptons and jets from {\tt Delphes} we list the selection cuts used in our $\ell^{\pm} \ell^{\pm} + 2 j + E_T^{\text{miss}}$ study.
\begin{enumerate}
  \item {\it Exactly same-sign dilepton + $\geq$ 2 jets}: We select exactly a pair of same-sign dilepton with additional criteria that they must be separated by a distance $\Delta R_{\ell_1 \ell_2} > 0.3$ and must have an invariant mass $m_{\ell_1 \ell_2} > 20$ GeV. Electrons are required to be outside the calorimeter transition region ($1.37 <|\eta_e| < 1.52$). To avoid additional background contributions from electron charge mis-reconstruction in di-electron events, we restrict electrons within $|\eta_e| < 1.37$ for such events, and discard events with $|m_{e_1 e_2} - m_Z | < 15$ GeV. We then require at least two jets in the selected event with $p_{T_j} > 20$ GeV and $|\eta_{j}| < 4.5$.

  \item {\it VBF cuts}: As mentioned before, a VBF event is characterized by two high-$p_T$ forward jets with large invariant mass and large separation in rapidity. Our signal is strictly produced by same-sign $W$ fusion along with the $W^{\pm} W^{\pm}jj$ (EW) background. In contrast, the di-jet invariant mass of QCD backgrounds peaks at smaller values and they are not widely separated in $|\Delta y_{j_1 j_2}|$, as can be seen in the lower left and right panels of Fig.~\ref{fig:distributions}. In contrast, the $p_T$ distributions of the leading jet $j_1$ and the sub-leading jet $j_2$ for the SM backgrounds tend to be flatter than those for the signal, as shown in the top left and right panels of Fig.~\ref{fig:distributions}.  We impose $p_{T_{j_1}} > 65$ GeV, $p_{T_{j_2}} > 35$ GeV, $m_{j_1 j_2} > 500$ GeV, and  $|\Delta y_{j_1 j_2}| > 2$ to select VBF topology. On an interesting note, although both the signal and $W^{\pm} W^{\pm}jj$ (EW) background are predominantly produced by same-sign $W$ fusion, their $|\Delta y_{j_1 j_2}|$ distributions do not peak at the same value. We attribute this difference to the contamination of $W^{\pm} W^{\pm}jj$ (EW) production by non-VBF processes. On the other hand, the signal production is strictly VBF.


  \item {\it $b$-jet veto}: Although we do not simulate same-sign dilepton backgrounds arising from heavy-flavor hadron decays, we include $b$-veto in our cut strategy following Ref.~\cite{Aaboud:2019nmv} to suppress such backgrounds.

  \item {$E_T^{\text{miss}}$ cut}: As the leptonic scalar $\phi$ decays invisibly into neutrinos, the missing transverse energy $E_T^{\text{miss}}$ tends to be slightly larger in the signal than in backgrounds. The $E_T^{\text{miss}}$ distributions for the signal and backgrounds are shown in the lower left panel of Fig.~\ref{fig:distributions1}, before VBF cuts. We impose a nominal $E_T^{\text{miss}}$ cut of 30 GeV.

  \item {\it Lepton $p_T$ cuts}: In the SM, the charged leptons from the $W$ boson decay have a peak at around $\sim m_W/2$. On the other hand, if a $\phi$ scalar is produced from a $W$ fusion process, it tends to be soft and most of the energy of the system is more likely to be carried by the leptons. As a result, the signal lepton $p_T$ distribution is much flatter and peaks around $\sim 2 m_W$, as can be seen in the top panels of Fig.~\ref{fig:distributions1}. We employ $p_{T_{\ell_1}} > 150$ GeV and $p_{T_{\ell_2}} > 90$ GeV to reduce $W^{\pm} W^{\pm}jj$ (EW) and  $W^{\pm} Zjj$ backgrounds by an order of magnitude.

  \item {\it $|\Delta \phi_{\ell_1,E_T^{\text{miss}} }|$ cut}: Finally, we use  $|\Delta \phi_{\ell_1,E_T^{\text{miss}} }| > 1.8$ to enhance the signal-to-background ratio, and that leads to a signal yield comparable to both $W^{\pm} W^{\pm}jj$ (EW) and  $W^{\pm} Zjj$ backgrounds. This cut is very effective due to different origins of $E_T^{\text{miss}}$ in the signal and the $W^{\pm} W^{\pm}jj$ (EW) background. While in the above background both the leptons and $E_T^{\text{miss}}$ are coming from $W$ boson decays, for the signal the $E_T^{\text{miss}}$ is arising from $\phi$ decay and leptons are emitted by incoming virtual $W$ bosons, which leads to different azimuthal angle correlation between them. This cut is very effective in suppressing the $W^{\pm} Zjj$ background as well. The $|\Delta \phi_{\ell_1,E_T^{\text{miss}} }|$ distributions are shown in the lower right panel of Fig.~\ref{fig:distributions1}.

\end{enumerate}

The cross sections after each set of cuts are shown in Table~\ref{tab:cutflow} for both the signal and SM backgrounds we considered. We show the cut-flow for $m_{\phi} =1 $ GeV case  with the couplings $|\lambda_{\alpha,\beta}| = 1\, (\alpha, \beta = e, \mu)$. The contribution from QCD $W^{\pm} W^{\pm} jj$ is negligible after all the cuts. However, both EW $W^{\pm} W^{\pm} jj$ and $W^{\pm} Z jj$ survive,  with background rates comparable to the expected signal rate when we apply the specific cuts $p_{T_{\ell_1}} > 150$ GeV, $p_{T_{\ell_2}}> 90$ GeV and $|\Delta \phi_{\ell_1,E_T^{\text{miss}} }| > 1.8$ in Table~\ref{tab:cutflow}. We repeat the above analysis for a number of benchmark points with $m_{\phi}$ in the range [100 MeV, 246 GeV]. Since we are selecting a highly boosted system by using VBF topological cuts, the cut-efficiencies of all the cuts shown in Table~\ref{tab:cutflow} show a weak dependence on $m_{\phi}$.
It is interesting to note that even if the scalar mass $m_\phi < 100$ MeV, the production cross sections and the $\lambda_{\alpha\beta}$ sensitivities will remain unchanged. For such light scalars there is no other direct limit from the LHC, although the low-energy high-precision constraints are much more stringent.
We do not explore $\phi$ masses beyond the EW scale of $v \simeq 246$ GeV since the effective Lagrangian of Eq.~(\ref{eqn:Lagrangian}) may not be valid in that regime.

\begin{table}[!tp]
\begin{center}
\caption{\label{tab:cutflow} Cut-flow table of the signal, with $m_\phi = 1$ GeV, and SM backgrounds $W^\pm W^\pm jj$ (EW), $W^\pm W^\pm jj$ (QCD) and $W^\pm Z jj$ at 14 TeV LHC. We decay $W^{\pm} \, (Z)$ boson to $\ell^{\pm} \nu \, (\ell^+ \ell^-)$, where $\ell = e, \mu, \tau$ during generation. In contrast, for the signal only $\ell = e, \mu$ are considered. The couplings $|\lambda_{\alpha,\beta}| \, (\alpha, \beta = e, \mu)$ are set to be 1. Note that the particular cuts in the last two rows can suppress very effectively the SM backgrounds.}
\resizebox{\columnwidth}{!}{%
\begin{tabular}{|c|c|c c c|}
\hline \hline
\multirow{2}{*}{\bf Cut selection} & {\bf Signal} & $W^{\pm} W^{\pm} jj$ (EW) & $W^{\pm} W^{\pm} jj$ (QCD) & $W^{\pm} Z jj$ \\
 & [fb] & [fb] & [fb] & [fb] \\
\hline
\hline
Production                  & 0.782 & 39.0 & 34.5 & 594  \\
\hline
  \tabincell{c}{ exactly $2\ell$: \\  $p_{T_{\ell_{1,2}}} > 10$ GeV, $|\eta_{\ell_{1,2}}|<2.5$, \\  $m_{\ell_1 \ell_2} > 20$ GeV, $\Delta R_{\ell_1 \ell_2} > 0.3$} & 0.530 & 9.26 & 5.65 & 177 \\
\hline
same-sign dilepton          & 0.529 & 9.26 & 5.65 &  44.5 \\
\hline
\tabincell{c}{for di-electron events: $|\eta_{e_1,e_2}|>1.37$, \\ $|m_{e_1e_2} -m_Z| <15$ GeV vetoed}
                            & 0.476 & 7.90 & 4.71 & 36.5   \\
\hline

 \tabincell{c}{$\geq 2$ jets: \\ $p_{T_{j_{1,2}}}  > 20$ GeV, $|\eta(j_{1,2})|<4.5$}
                            & 0.397 & 7.46 & 4.51 & 33.7 \\
\hline
 \tabincell{c}{VBF cuts: \\ \small $p_{T_{j_1}} > 65$ GeV, $p_{T_{j_2}} > 35$ GeV, \\ \small $m_{j_1 j_2} > 500$ GeV, $|\Delta y_{j_1 j_2}| > 2$}
                            & 0.165 & 4.08 & 0.502 & 3.42 \\
\hline
 $b$-jet veto               & 0.158 & 3.77 & 0.441 & 3.03 \\
 \hline
 $E_T^{\text{miss}} > 30$ GeV
                            & 0.143 & 3.41 & 0.399 & 2.58 \\
\hline\hline
 {$p_{T_{\ell_1}} > 150$ GeV, $p_{T_{\ell_2}}> 90$ GeV}
                            & 0.108 & 0.217 & 0.017 & 0.176 \\
\hline
$|\Delta \phi_{\ell_1,E_T^{\text{miss}} }| > 1.8$
                            & 0.084 & 0.088 & 0.004 & 0.059                            \\
\hline \hline
\end{tabular}}
\end{center}
\end{table}

In Table~\ref{tab:evtyield} we present event yields in different lepton flavor combinations $e^\pm e^\pm$, $e^\pm \mu^\pm$ and $\mu^\pm \mu^\pm$ for both the signal (with $m_{\phi} =1 $ GeV) and  SM backgrounds at 14 TeV LHC with 3 ab$^{-1}$ of integrated luminosity. As we  mentioned before,  $|\lambda_{\alpha,\beta}|\, (\alpha, \beta = e, \mu)$ are set to be 1. If we switch on couplings involving $\tau$ leptons as well we can get $\sim 15 \%$ enhancement on the signal yield. In Table~\ref{tab:evtyield} we also calculate the significance of the signal in different channels for $0\%$ and $10\%$ systematic errors on the background estimation, using the metric $\sigma_{S/B} = {S}/{\sqrt{S+B+(\epsilon_B B)^2}}$. Here $\epsilon_B$ is the percentage systematic error on the background estimation.

\begin{table}[!tp]
\begin{center}
\caption{\label{tab:evtyield} Event yields in different lepton flavor combination channels $e^\pm e^\pm$, $e^\pm \mu^\pm$ and $\mu^\pm \mu^\pm$ for both the signal and SM backgrounds at 14 TeV LHC with 3 ab$^{-1}$ of integrated luminosity. For the signal we set $m_{\phi} =1$ GeV and $|\lambda_{\alpha,\beta}|=1$ (with $\alpha, \beta = e, \mu$). We consider systematic errors of 0\%, 10\% and 20\% on the background events only.}
\vspace{0.2cm}
\begin{tabular}{|c|c| c c c|c|}
\hline \hline
\multicolumn{2}{|c|}{\bf Channels} & $e^{\pm} e^{\pm}$ & $e^{\pm} \mu^{\pm}$ & $\mu^{\pm} \mu^{\pm}$ & {\bf Total} \\ \hline\hline
\multicolumn{2}{|c|}{Signal}            & 40 & 129 & 84 & 253 \\ \hline
\multicolumn{2}{|c|}{$W^{\pm} W^{\pm} jj$ (EW)}         & 37 & 137 & 89 & 263 \\
\multicolumn{2}{|c|}{$W^{\pm} W^{\pm} jj$ (QCD)}          & 2 & 9 & 2 & 13 \\
\multicolumn{2}{|c|}{$W^{\pm} Z jj$}         & 29 & 94 & 54 & 177 \\ \hline
\multicolumn{2}{|c|}{Total background}   & 68 & 240 & 145 & 453 \\ \hline\hline
\multirow{3}{*}{Significance} & syst. error $0\%$ & 3.87 & 6.73 & 5.53 & 9.53 \\
 & syst. error $10\%$ & 3.24 & 4.21 & 4.00 & 4.83 \\
 & syst. error $20\%$ & 2.35 & 2.50 & 2.56 & 2.68 \\
 \hline \hline
\end{tabular}
\end{center}
\end{table}

\subsection{Prospects} \label{sec:prospects}

\begin{figure}
  \centering
  \includegraphics[height=0.5\textwidth]{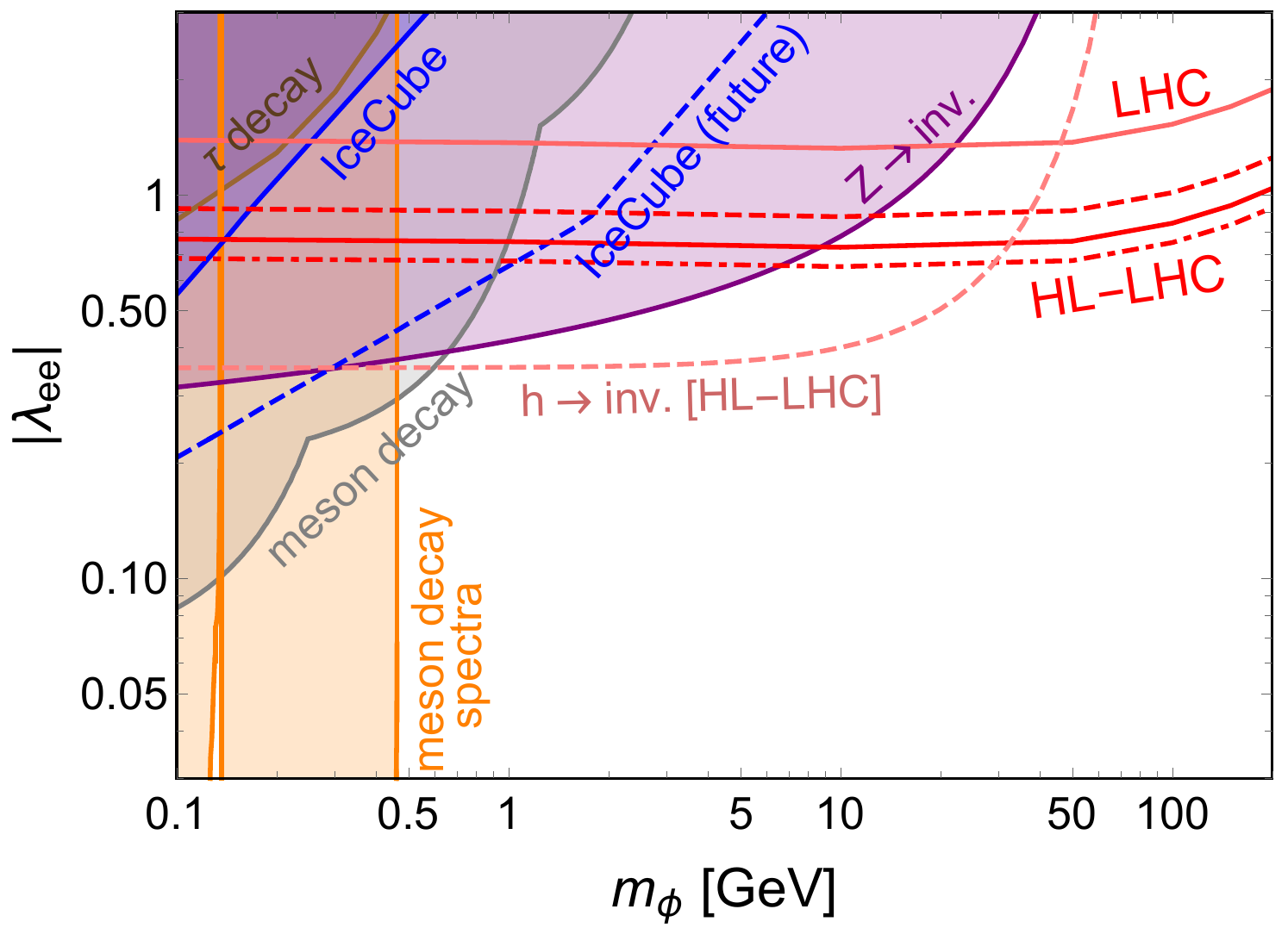}
  \caption{Prospects of the coupling $|\lambda_{ee}|$ as a function of the scalar mass $m_\phi$ at 14 TeV LHC with luminosity of 300 fb$^{-1}$ (solid thin red line) and HL-LHC with 3 ab$^{-1}$ and with systematic errors of 0\% (dot-dashed thick red line),  10\% (solid thick red line) and 20\% (dashed thick red line). Also shown are the low-energy limits (cf.~Table~\ref{tab:limits}) from meson decay (gray), $\tau$ decay (brown), heavy neutrino searches in meson decay spectra (orange), invisible $Z$ decay (purple) and the prospect of invisible SM Higgs decay at HL-LHC (dashed pink), the current IceCube limits on neutrino--neutrino interactions (blue) and prospects (dashed blue). All the shaded regions are excluded.}
  \label{fig:LHC}
\end{figure}

\begin{figure}
  \centering
  \includegraphics[height=0.5\textwidth]{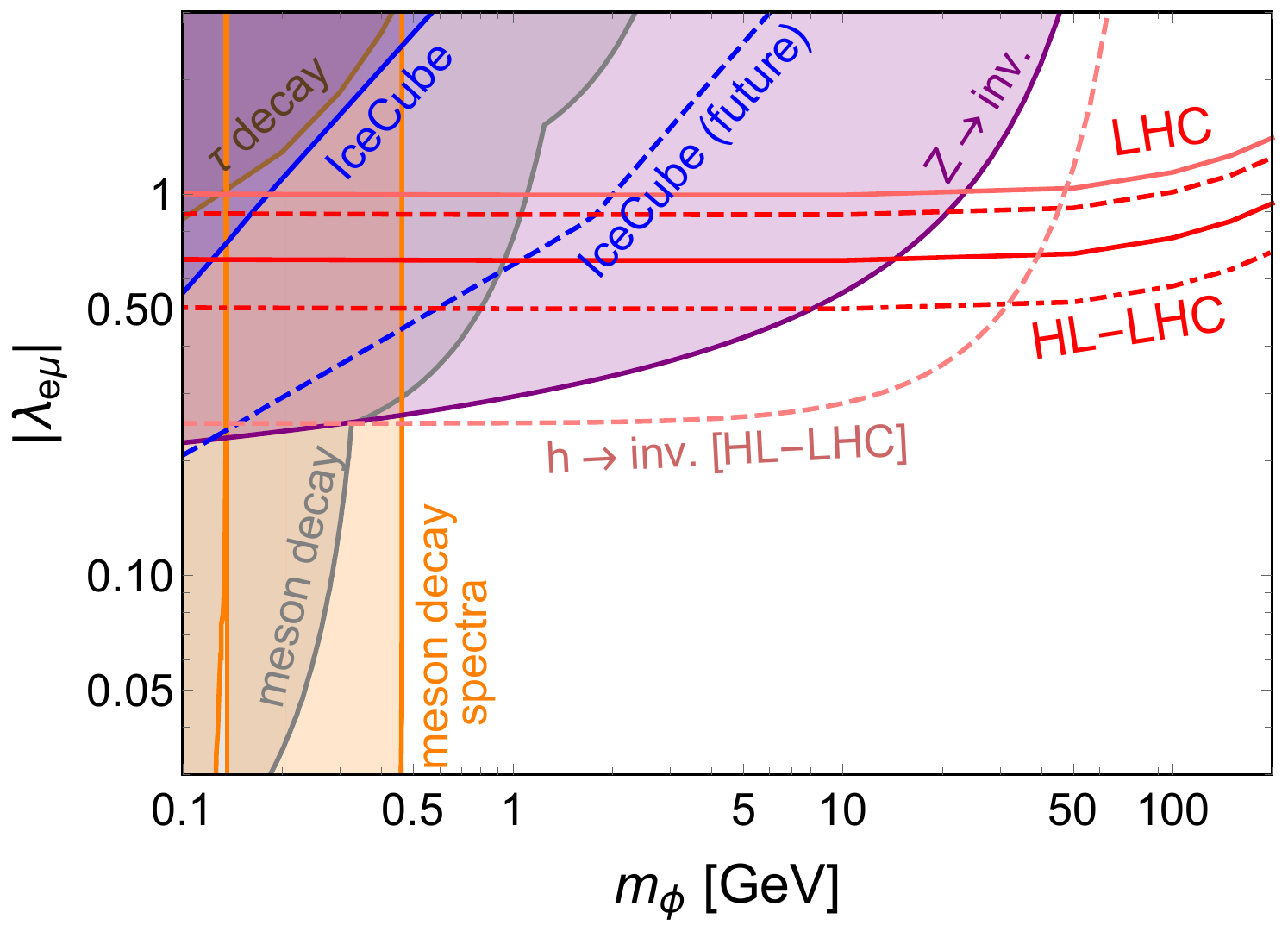}
  \caption{The same as in Fig.~\ref{fig:LHC}, but for the coupling $|\lambda_{e\mu}|$.}
  \label{fig:LHC2}
\end{figure}

\begin{figure}
  \centering
  \includegraphics[height=0.5\textwidth]{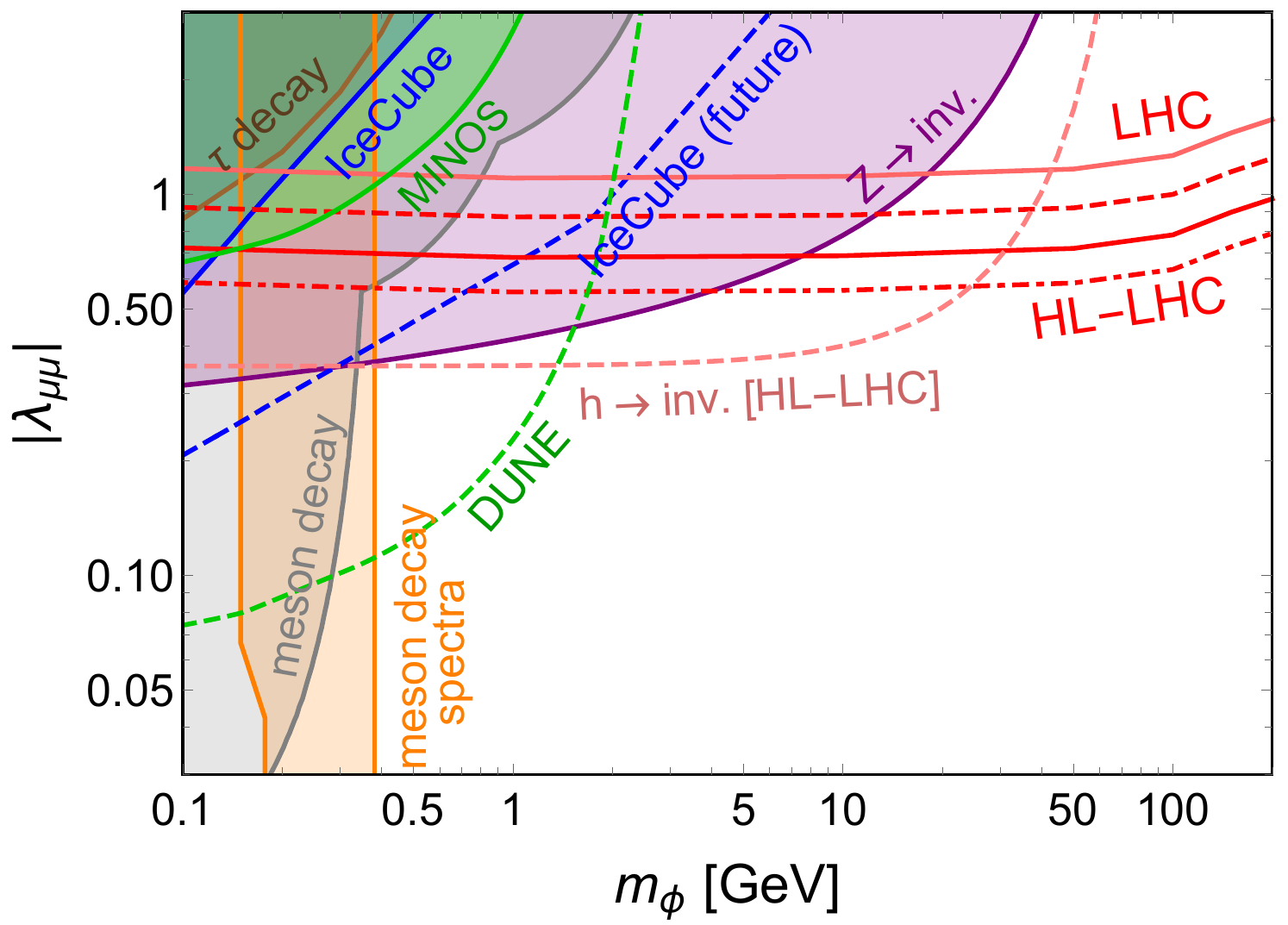}
  \caption{The same as in Fig.~\ref{fig:LHC}, but for the coupling $|\lambda_{\mu\mu}|$. Here we also show the limit on $|\lambda_{\mu\mu}|$ from MINOS (green) and prospect at DUNE (dashed green).}
  \label{fig:LHC3}
\end{figure}

\begin{table}[!tp]
\begin{center}
\caption{\label{tab:results} Summary of the 95\% C.L. LHC and HL-LHC sensitivities to the couplings $|\lambda_{\alpha\beta}|$ in our leptonic scalar case with $m_\phi\lesssim 50$ GeV [cf.~Figs.~\ref{fig:LHC}--\ref{fig:LHC3}]. Results with 0\%, 10\% and 20\% systematic errors are listed.}
\vspace{0.2cm}
\begin{tabular}{|c|c| c c c|}
\hline \hline
\multicolumn{2}{|c|}{\bf Collider} & $|\lambda_{ee}|$ & $|\lambda_{e\mu}|$ & $|\lambda_{\mu\mu}|$ \\ \hline\hline
\multirow{3}{*}{LHC} &  syst. error $0\%$ & 1.35 & 0.95 & 1.07 \\
& syst. error $10\%$ & 1.38 & 1.00 & 1.13 \\
& syst. error $20\%$ & 1.42 & 1.09 & 1.19 \\
\hline
\multirow{3}{*}{HL-LHC} &  syst. error $0\%$ & 0.68 & 0.51 & 0.57 \\
& syst. error $10\%$ & 0.76 & 0.68 & 0.70
 \\
 & syst. error $20\%$ & 0.91 & 0.88 & 0.87 \\
 \hline \hline
\end{tabular}
\end{center}
\end{table}

The prospects of $\lambda_{ee,\,e\mu,\, \mu\mu}$ at the LHC and HL-LHC are shown respectively in Figs.~\ref{fig:LHC}, \ref{fig:LHC2} and \ref{fig:LHC3}.
The dot-dashed thick red lines are for the most optimistic case at the 14 TeV HL-LHC with 3 ab$^{-1}$ integrated luminosity and without any systematic error, where the couplings $\lambda_{ee,\,e\mu,\, \mu\mu}$ can be probed respectively up to 0.68, 0.51 and 0.57 at the 95\% C.L (see Table~\ref{tab:results}). With a realistic 10\% (20\%) systematic error, the sensitivities at the HL-LHC are slightly weaker, being respectively 0.76 (0.91), 0.68 (0.90) and 0.70 (0.89) at the 95\% C.L., denoted by the solid (dashed) thick red lines.
This implies that our leptonic signals are rather robust against the systematic uncertainties on the background determination.
For comparison,  we also show the prospects at the 14 TeV LHC  with only 300 fb$^{-1}$ integrated luminosity, which is achievable in the upcoming run within a few years. We use the same cuts above as for the HL-LHC and assume there is a 10\% (20\%) systematic error. The prospects are respectively 1.38 (1.42), 1.00 (1.09), and 1.13 (1.19)  at the 95\% C.L. for the couplings $\lambda_{ee,\,e\mu,\, \mu\mu}$. The corresponding LHC prospects with zero systematic uncertainty are respectively 1.35, 0.95 and 1.07, as shown in Table~\ref{tab:results}. Since the difference between the LHC prospects with 0\%, 10\% and 20\% systematic uncertainties is not appreciable, we show only the prospects with 10\% systematic error as the thin red lines in Figs.~\ref{fig:LHC}--\ref{fig:LHC3}.
The slightly better  sensitivity for $\lambda_{e\mu}$ is due to the doubling of the flavor combinations; see the event rates with different lepton flavors estimated in Table  \ref{tab:evtyield}.
We find that when the scalar mass is significantly smaller than the colliding energy at LHC, say $m_\phi \lesssim 50$ GeV, the sensitivities have only a weak dependence on the scalar mass, being almost flat.
We also note that although the production cross section of $\phi$ at the LHC via VBF process starts falling for $m_{\phi} \gtrsim 10$ GeV as can be seen from Fig.~\ref{fig:prod_xsec}, the sensitivity curves slowly drop when the scalar mass $m_\phi \gtrsim 50$ GeV only. For the $\phi$ mass between 10 and 50 GeV the small decrease in the production cross section is compensated by similar increase in cut-efficiencies. Above 50 GeV, improvements in cut-efficiencies can not overcome the sharp drop in cross sections.


The prospects of $\lambda_{\alpha\beta}$ at the LHC and HL-LHC are largely complementary to the low-energy constraints discussed in section~\ref{sec:limits}. To see it more clearly, we show in Figs.~\ref{fig:LHC}--\ref{fig:LHC3} the limits from meson decays (gray), $\tau$ decays (brown), heavy neutrino searches in two-body meson decay spectra (orange), the invisible $Z$ decay (purple), neutrino-matter scattering at MINOS (green),
and IceCube limits on new neutrino--neutrino interactions (blue). As in Figs.~\ref{fig:meson}--\ref{fig:Zinv}, all the shaded regions are excluded. Also shown are the prospects of invisible SM Higgs decay at HL-LHC by the dashed pink lines and the prospects at IceCube-Gen2 by the dashed blue lines. For the coupling $\lambda_{\mu\mu}$ we have also shown the prospect from DUNE in Fig.~\ref{fig:LHC3} by dashed green line. One can see from Figs.~\ref{fig:LHC}--\ref{fig:LHC3} that the HL-LHC prospects of $\lambda_{ee,\,e\mu,\, \mu\mu}$ exceed all the existing limits when the scalar mass $m_\phi \gtrsim 10$ GeV.  These will be the first direct collider limits on light leptonic scalars of such kind.

\section{Conclusion}
\label{sec:conclusion}

In this paper, we have studied the neutrino non-standard interaction in a simplified framework where a (light) scalar $\phi$ couples exclusively to the active neutrinos. As such, it carries two units of lepton-number, and is dubbed as ``leptonic scalar''. The Yukawa couplings $\lambda_{\alpha\beta}$ of $\phi$ to neutrinos are constrained by different low-energy, high-precision data, such as the charged meson and lepton decays rates, meson decay spectra, the $W$, $Z$ and $h$ decay rates, neutrino beam experiments like MINOS and, in the future, DUNE,
IceCube and CMB limits on new neutrino self interactions, as well as other limits which are mostly relevant to a light scalar with mass $m_\phi \lesssim 100$ MeV.

We have shown that the leptonic scalar $\phi$ can be produced at high-energy hadron colliders like the LHC via fusion of two same-sign $W$ bosons, i.e., $pp \to \ell_{\alpha}^\pm \ell_{\beta}^\pm jj + \phi$. As $\phi$ decays into neutrinos, we have the distinctive signature $pp \to \ell_\alpha^\pm \ell_\beta^\pm jj + E_T^{\rm miss}$. The predominant SM background is the electroweak vector boson scattering process $pp \to W^\pm W^\pm jj$, with sub-leading contributions from the $W^\pm Zjj$ production and QCD $W^\pm W^\pm jj$ processes. Given the presence of the scalar $\phi$, the kinematic distributions of the charged leptons, jets and missing transverse energy are very different for the signal and backgrounds and can be used to effectively separate the two. Upon dedicated simulations of both signal and backgrounds,
we find that for $m_{\phi}$ less than the electroweak scale, the couplings $\lambda_{ee,\, e\mu,\, \mu\mu}$ can be probed, respectively, up to 1.38, 1.00 and 1.13 at the 95\% C.L. at the LHC with an integrated luminosity of 300 fb$^{-1}$, and down to 0.76, 0.68 and 0.70 at the HL-LHC with 3 ab$^{-1}$ integrated luminosity; see Table~\ref{tab:results} for a quick summary. Based on this analysis, we find that the direct constraints from the ongoing LHC would be better than all other existing constraints for $m_\phi\gtrsim 10$ GeV. Figs.~\ref{fig:LHC}, \ref{fig:LHC2} and \ref{fig:LHC3} summarize our results for the couplings $|\lambda_{\alpha\beta}|$ with $\alpha\beta=ee,e\mu,\mu\mu$ respectively. At future colliders, such as a high-energy upgrade of the LHC or a future 100 TeV $pp$ collider, an improved sensitivity is expected. In some sense, this is a direct probe of scalar-mediated neutrino self-interactions at high-energy colliders, and is largely complementary to the constraints from the low-energy high-intensity experiments and the astrophysical and cosmological observations.

\section*{Acknowledgements}
We thank Kevin Kelly for useful discussion. This work was initiated at the NTN NSI workshop, supported in part by the US Neutrino Theory Network Program under Grant No. DE-AC02-07CH11359, as well as by the Department  of  Physics  and  the  McDonnell  Center  for  the  Space  Sciences  at  Washington University in St.  Louis. PSBD would like to thank the Fermilab Theory Group for warm hospitality where part of this work was done. YZ would like to thank the Institute of Theoretical Physics, Chinese Academy of Sciences, the Tsung-Dao Lee Institute, and the Institute of High Energy Physics, Chinese Academy of Sciences for generous hospitality where part of this work was done.
The work of AdG was supported in part by the U.S.~Department of Energy under grant No. DE-SC0010143.
The work of PSBD and YZ was supported in part by the U.S.~Department of Energy under Grant No. DE-SC0017987 and in part by the MCSS.
The work of BD was supported in part by the U.S.~Department of Energy under grant No. DE-SC0010813.
TG was supported by U.S.~Department of Energy under grant No. DE-SC0010504.
The work of TH was supported in part by the U.S.~Department of Energy under grant No.~DE-FG02-95ER40896 and in part by the PITT PACC.
This work was partially performed at the Aspen Center for Physics, which is supported by National Science Foundation grant PHY-1607611.

\appendix

\section{Possible UV Complete Models for Leptonic Scalar}
\label{app:UV}

In our analysis, we have introduced the neutrino coupling to the leptonic scalar $\phi$ via an effective dimension-six operator $(LH)(LH)\phi/\Lambda^2$~\cite{Berryman:2018ogk,Kelly:2019wow} that gives rise to the $\nu\nu\phi$ interaction in Eq.~\eqref{eqn:Lagrangian}. In this appendix, we discuss several possible UV-complete models that, after integrating out the heavy degrees of freedom, lead to this effective operator. Most of the models discussed here are inspired by the tree-level seesaw realizations of the dimension-five, Weinberg operator $(LH)(LH)/\Lambda$~\cite{Weinberg:1979sa}, except that all new particles introduced here preserve the $B-L$ symmetry. The specific details of these model frameworks are irrelevant for the model-independent collider analysis performed in the main text, assuming that the new particles introduced in these models are much heavier than the leptonic scalar $\phi$.

One option is to introduce an $SU(2)_L$-triplet scalar $\Delta$ with hypercharge $+1$ and $B-L$ charge $+2$. The relevant renormalizable Lagrangian in this case is given by (see also Refs.~\cite{Berryman:2018ogk, Dey:2018uvu})
\begin{align}
{\cal L} \ \supset \ y_{\alpha\beta} L_\alpha \Delta L_\beta +\lambda_\Delta \phi H \Delta^\dag H -M^2_\Delta {\rm Tr}(\Delta^\dag \Delta)+{\rm H.c.}
\end{align}
This is similar to the type-II seesaw model~\cite{Schechter:1980gr, Cheng:1980qt, Mohapatra:1980yp, Lazarides:1980nt}, but there are no $B-L$ violating terms here.
Once the $\Delta$-field is integrated out, the above-mentioned dimension-six operator is produced, with the effective $\lambda$-couplings in Eq.~\eqref{eqn:Lagrangian} identified as
\begin{align}
\lambda_{\alpha\beta} \ = \ \frac{y_{\alpha\beta}\lambda_\Delta v^2}{M_\Delta^2} \, ,
\end{align}
which can be large, provided that the mass of the $\Delta$-field is close to the EW scale, depending on the flavor structure. This $\Delta$ field, and in particular, its doubly-charged component $\Delta^{\pm\pm}$ offers its own rich collider phenomenology~\cite{Perez:2008ha, Melfo:2011nx, Fuks:2019clu}. There are stringent constraints on the lepton Yukawa couplings $y_{\alpha\beta}$ from collider searches, as well as searches for low-energy lepton flavor violating processes~\cite{Dev:2018kpa}, however, $\lambda_{\alpha\beta}\sim {\cal O}(1)$, as required to be relevant for the LHC sensitivity study, is still achievable in this case.

Another option is to introduce pairs of vector-like fermions $N_i$ and $N_i^c$ (with $i=1,2,\cdots, n$) which are SM singlets with $B-L$ charges $\mp 1$, respectively. The relevant renormalizable Lagrangian is given by
\begin{align}
    {\cal L} \ \supset \ y_{\alpha i}L_{\alpha} H N_i^c+\lambda_{N,ij}\phi N_i N_j+M_{N,i}N_i N_i^c+{\rm H.c.}
    \label{eq:A3}
\end{align}
This is similar to the type-I seesaw model~\cite{Minkowski:1977sc, Mohapatra:1979ia, Yanagida:1979as, GellMann:1980vs}, but there are no $B-L$ violating terms here. After integrating out the heavy vector-like fermion fields, we obtain the desired  dimension-six operator, with the effective $\lambda$-couplings in Eq.~\eqref{eqn:Lagrangian} given by
\begin{align}
    \lambda_{\alpha\beta} \ = \ y_{\alpha i}M_{N,i}^{-1}\lambda_{N,ij}M_{N,j}^{-1}y^T_{j\beta} \, .
    \label{eq:A4}
\end{align}
Here the Yukawa couplings $y_{\alpha i}$ also lead to the mixing of the SM neutrinos with the new vectorlike fermions, with the mixing angle $\theta\sim yv/M_N$, which is constrained to be $\lesssim {\cal O}(0.01)$ for $M_N>v$ from electroweak precision data~\cite{delAguila:2008pw, Antusch:2014woa, Flieger:2019eor}. Thus, the $\lambda$-couplings in Eq.~\eqref{eq:A4} cannot be of ${\cal O}(1)$ in this setup.

One can replace the SM-singlet vector-like fermions in Eq.~\eqref{eq:A3} by $SU(2)_L$-triplet fermions, as in the type-III seesaw model~\cite{Foot:1988aq}. In this case, the Yukawa couplings will be of the form $y_{\alpha i}L_\alpha \sigma^a H N_{ia}$, where $a=1,2,3$ is the $SU(2)_L$ index in the adjoint representation and $\sigma^a$ are the Pauli matrices. After integrating out the heavy $N_i$ fields, the low-energy effective operator takes the form $(L\sigma^a H)(L\sigma_a H)\phi/\Lambda^2$, with the effective $\phi\nu\nu$ coupling related to the UV parameters in the same way as in Eq.~\eqref{eq:A4}. Nonetheless, the experimental constraints on $y$ are still applicable in this case, thus ruling out the possibility of large $\lambda_{\alpha\beta}$.

Similar examples of UV-complete models for Majorana neutrinos were discussed recently in Ref.~\cite{Blinov:2019gcj}. Using the scalar field as a portal to the dark sector was discussed in Ref.~\cite{Kelly:2019wow}, where an additional $Z_2$, $Z_3$ or $U(1)$ symmetry was invoked to stabilize DM.

The effective coupling of $\phi$ to neutrinos as in Eq.~\eqref{eqn:Lagrangian} is similar to that of a Majoron~\cite{Chikashige:1980qk, Chikashige:1980ui, Aulakh:1982yn, Gelmini:1980re, Schechter:1981cv} -- the (pseudo) Goldstone boson from spontanesouly broken lepton number. The equivalent coupling $\lambda$ in this case is related to the observed neutrino masses, $\lambda \sim m_\nu/f$, where $f$ is the spontaneous lepton number breaking scale. Since we are mostly interested in sizable couplings $\lambda\sim {\cal O}(1)$, the lepton number breaking scale would have to be very low, $f\sim m_\nu\lesssim {\cal O}(1~{\rm eV})$.

Another possibility to explain the leptophilic nature of the $\phi$ field exists in the framework of the left-right symmetric model (LRSM)~\cite{LR1, LR2, LR3}. In this case, $\phi$ can be identified as the neutral component of the $SU(2)_R$-triplet $\Delta_R$, which can be below the EW scale in some region of the parameter space~\cite{Dev:2017dui}, depending on its corresponding scalar quartic coupling and radiative corrections. Its coupling to the SM neutrinos arises from its direct coupling to the heavy right-handed neutrinos, which then mix with the light active neutrinos, with the mixing angle again constrained to be at most ${\cal O}(0.01)$ (see e.g. Ref.~\cite{Bolton:2019pcu}).

Yet another scenario to naturally explain the neutrinophilic nature of the leptonic scalar is in the context of a neutrinophilic two Higgs doublet model~\cite{Wang:2006jy, Gabriel:2006ns}, where one of the Higgs doublets has a very small VEV, of ${\cal O}({\rm eV})$, and is responsible for the tiny neutrino masses. The neutral component of this second Higgs doublet can be identified as our leptonic scalar $\phi$. However, the astrophysical and cosmological constraints severely restrict the neutrino Yukawa coupling to be $\lesssim 10^{-5}$ in this case~\cite{Sher:2011mx, Zhou:2011rc}, and thus rule out the possibility of having $\lambda\sim {\cal O}(1)$. In addition, the effective coupling of the scalar $\phi$ with neutrinos in a neutrinophilic doublet model will be $\nu \bar{\nu} \phi$, instead of $\nu \nu \phi$. Hence, the relevant final state to study at the LHC will be $\ell^+_\alpha \ell^-_\beta \phi \, + 2j$, which will suffer from significantly large SM backgrounds coming from $W^+ W^-+$jets and $t \bar{t}+$jets.

\section{Calculations of Multi-body Decays Involving $\phi$}
\label{sec:decay}


\subsection{Three-body decays $Z \to \nu_\alpha \nu_\beta \phi$, $W \to \ell_\alpha \nu_\beta \phi$ and $h \to \nu_\alpha \nu_\beta \phi$}
\label{sec:decay1}

For the decay $Z \to \nu_\alpha (p_2) + \nu_\beta (p_3) + \phi (p_1)$ with the flavor index $\alpha = e,\,\mu,\,\tau$ and $p_{1,2,3}$ the momenta of particles in the final state, the scalar $\phi$ can be emitted from either of two neutrino lines, and the partial width reads
\begin{eqnarray}
\label{eqn:Zdecay}
\Gamma (Z \to \nu_\alpha \nu_\beta \phi) \ = \
\frac{g^2 |\lambda_{\alpha\beta}|^2}{12 m_Z (1+\delta_{\alpha\beta}) \cos^2\theta_W} \int {\rm d} \Phi_3
\left( |{\cal M}_1|^2 + |{\cal M}_2|^2 \right) \,,
\end{eqnarray}
with $m_Z$ the $Z$ boson mass, $\theta_W$ the weak mixing angle, $g$ the coupling constant for the SM gauge group $SU(2)_L$, and the two reduced amplitudes squared are respectively
\begin{eqnarray}
|{\cal M}_{1,2}|^2 & \ = \ &
\frac{1}{(p_1 + p_{3,2})^2} \bigg[
4 (p_1 \cdot p_2) (p_1 \cdot p_3) - 2 m_Z^2 x_{\phi Z} (p_2 \cdot p_3) \nonumber \\
&& \qquad \qquad \qquad + \frac{(p_2 \cdot p_3)}{m_Z^2} \left( m_Z^2 x_{\phi Z}^2 + 2 (p_1 \cdot p_{3,2}) \right)^2 \bigg] \,,
\end{eqnarray}
with $x_{\phi Z} = m_\phi^2/m_Z^2$, and the three-body phase space~\cite{Asatrian:2012tp}
\begin{eqnarray}
\nonumber {\rm d} \Phi_3 & \ = \ &
\frac{m_Z^{2} \Omega_{2}\Omega_{3}}{2^{10} \pi^5}
\frac{\lambda_2 (1-\lambda_2) (1-x_{\phi Z})^3}{\lambda_2 (1-x_{\phi Z})+x_{\phi Z}}
\,{\rm d}\lambda_1 {\rm d}\lambda_2 \,,
\end{eqnarray}
where $\Omega_{d} \equiv 2\pi^{d/2}/\Gamma(d/2)$ is the solid angle in $d$ dimension, and $0 < \lambda_{1,2} < 1$ are the two dimensionless kinematic variables. The scalar products of momenta can be expressed as functions of $m_{Z}$, $x_{\phi Z}$ and $\lambda_{1,2}$ in the following form~\cite{Asatrian:2012tp}:
\begin{eqnarray}
(p_1 \cdot p_2) & \ = \ & \frac{m_Z^2}{2}
\frac{(1-\lambda_1) (1-\lambda_2) [x_{\phi Z}+(1-x_{\phi Z}) \lambda_1 \lambda_2]}{\lambda_2 (1-x_{\phi Z}) + x_{\phi Z}} \,, \\
(p_2 \cdot p_3) & \ = \ & \frac{m_Z^2}{2}
\frac{\lambda_2 (1-\lambda_1) (1-\lambda_2) (1-x_{\phi Z})^2}{\lambda_2 (1-x_{\phi Z}) + x_{\phi Z}} \,, \\
(p_1 \cdot p_3) & \ = \ & \frac{m_Z^2}{2}
(1-x_{\phi Z}) \lambda_2 \,.
\end{eqnarray}

The calculation of partial width for the $W$ boson decay $W \to \ell_\alpha \nu_\beta \phi$ is quite similar, for which we have only one diagram, and the partial width is
\begin{eqnarray}
\label{eqn:Wdecay}
\Gamma (W \to \ell_\alpha \nu_\beta \phi) \ = \
\frac{g^2 |\lambda_{\alpha\beta}|^2}{6 m_W} \int {\rm d} \Phi_3 (m_W)
\left|{\cal M}_1 (m_W, x_{\phi W}) \right|^2 \,,
\end{eqnarray}
with $x_{\phi W} \equiv m_\phi^2 /m_W^2$.

The dimension-5 coupling of the SM Higgs $h$ with the light scalar $\phi$ and neutrinos in Eq.~(\ref{eqn:Lagrangian2}) induces the decay $h \to \nu_\alpha \nu_\beta \phi$, with the partial width
\begin{eqnarray}
\Gamma (h \to \nu_\alpha \nu_\beta \phi) \ = \
\frac{4 |\lambda_{\alpha\beta}|^2}{m_h v_{}^2} \int {\rm d} \Phi_3 (m_h)
\left|{\cal M}_3 (m_h, x_{\phi h}) \right|^2 \,,
\label{eq:hdecay}
\end{eqnarray}
with $x_{\phi h} = m_\phi^2/m_h^2$ and $
\left|{\cal M}_3 \right|^2 \ = \
(p_2 \cdot p_3)$, with $p_2, p_3$ being the outgoing neutrino four-momenta.

\subsection{Four-body decays $\tau \to \ell \nu\nu \phi$} \label{sec:decay2}

For the four-body decays $\tau \to \ell_\alpha (p_2) + \nu_\beta (p_3) + \nu_\gamma (p_4) + \phi (p_1)$ with the flavor indices $\alpha = e,\,\mu$ and $\beta,\,\gamma = e,\, \mu,\, \tau$, the scalar $\phi$ could couple to the $\nu_\alpha$ and/or the $\nu_\tau$ lines, depending on the flavor indices of the coupling $\lambda_{\rho\sigma}$. In particular, when $\rho = \beta = \alpha$ and $\sigma = \gamma = \tau$ we have two diagrams, and only one for other cases. For simplicity, we neglect the charged lepton mass in the final state, and the partial width is
\begin{eqnarray}
\label{eqn:taudecay}
\Gamma (\tau \to \ell_\alpha \nu\nu\phi) \ \simeq \ \frac{32 G_F^2 |\lambda_{\beta\gamma}|^2}{m_\tau (1+\delta_{\rho \beta} \delta_{\sigma\tau})} \int {\rm d} \Phi_4
\left( \delta_{\rho \alpha} |{\cal M}_1|^2 + \delta_{\sigma \tau} |{\cal M}_2|^2 \right)  \,,
\end{eqnarray}
where the $\delta$ factors in the denominator account for identical neutrinos in the final state, and the reduced amplitudes squared are
\begin{eqnarray}
\left| {\cal M}_1 \right|^2 & \ = \ &
\frac{\left( p_2 \cdot p_4 \right)}{(p_1+p_3)^2}
\left[ 2\left( p_2 \cdot p_4 \right) p_1 \cdot (p_2+p_3+p_4) +
m_\tau^2 x_{\phi\tau} p_3 \cdot (p_1 - p_2 - p_4) \right] \,, \\
\left| {\cal M}_2 \right|^2 & \ = \ &
\frac{(p_1 \cdot p_3) + (p_2 \cdot p_3) + (p_3 \cdot p_4)}{(p_1+p_4)^2}
\left[ 2(p_1 \cdot p_2) (p_1 \cdot p_4) - m_\tau^2 x_{\phi \tau} (p_2 \cdot p_4) \right] \,,
\end{eqnarray}
with $x_{\phi\tau} \equiv m_\phi^2/m_\tau^2$, and the four-body phase space is~\cite{Asatrian:2012tp}
\begin{eqnarray}
\nonumber {\rm d} \Phi_4 & \ = \ &
\frac{m_\tau^{4}}{128} \,
\frac{\Omega_{1}\Omega_{2}\Omega_{3}}{(2\pi)^{8}}
\,{\rm d}\lambda_1 {\rm d}\lambda_2 {\rm d}\lambda_3
{\rm d}\lambda_4 {\rm d}\lambda_5 \\
&& \times (1-\sqrt{x_{\phi\tau}})^{5} \left(\lambda_1(1-\lambda_2)\right)
\left(\lambda_5(1-\lambda_5)\right)^{-1/2}  \nonumber \\
&& \times \left[(1-\lambda_1)((1+\sqrt{x_{\phi\tau}})^2-
\lambda_1(1-\sqrt{x_{\phi\tau}})^2)\right]^{1/2} \,,
\end{eqnarray}
where $0 < \lambda_{i} < 1$ (with $i = 1$ to $5$) are dimensionless kinematic variables. The scalar products of momenta can be expressed as functions of $m_{\tau}$, $x_{\phi\tau}$ and the $\lambda$'s as follows~\cite{Asatrian:2012tp}:
\begin{eqnarray}
\left( p_1 \cdot p_2 \right) & \ = \ &
E_2 \left(E_1 - \sqrt{E_1^2 - x_{\phi\tau} m_\tau^2} \cos\theta_1 \right) \,, \\
\left( p_1 \cdot p_3 \right) & \ = \ &
\frac12 \left[ (s_{13}^+-s_{13}^-)\lambda_5+s_{13}^- - m_\tau^2 x_{\phi\tau} \right] \,, \\
\left( p_1 \cdot p_4 \right) & \ = \ &
E_1 \sqrt{s_{234}} - \left( p_1 \cdot p_2 \right) - \left( p_1 \cdot p_3 \right) \,, \\
\left( p_2 \cdot p_3 \right) & \ = \ &
\frac12 m_\tau^2(1-\sqrt{x_{\phi\tau}})^2 \lambda_1
(1-\lambda_2) \lambda_4 \,, \\
\left( p_2 \cdot p_4 \right) & \ = \ &
E_2 \sqrt{s_{234}} - \left( p_2 \cdot p_3 \right) \,, \\
\left( p_3 \cdot p_4 \right) & \ = \ &
\frac12 m_\tau^2 (1-\sqrt{x_{\phi\tau}})^2 \lambda_1 \lambda_2 \,,
\end{eqnarray}
where
\begin{eqnarray}
E_1 & \ = \ & \frac{m_\tau^2 (1-x_{\phi\tau})-s_{234}}{2\sqrt{s_{234}}} \,,  \\
E_2 & \ = \ & \frac{1}{2}m_\tau \sqrt{\lambda_1}(1-\lambda_2) (1-\sqrt{x_{\phi\tau}}) \,, \\
s_{234} & \ = \ & m_\tau^2 (1-\sqrt{x_{\phi\tau}})^2 \lambda_1 \,, \\
\cos\theta_1 & \ = \ & 2\lambda_3-1 \,, \\
s_{13}^{\pm} & \ = \ & m_\tau^2 \left[ x_{\phi\tau} +\frac{1}{2}\,(1-\sqrt{x_{\phi\tau}})\left[\left(\lambda_2
\left(1-\lambda_4\right)+\lambda_4\right)
\left(1+\sqrt{x_{\phi\tau}}-\lambda_1(1-\sqrt{x_{\phi\tau}})\right)\right. \right.
\nonumber \\ &&
+\left.\left(\lambda_2(1-\lambda_4)-\lambda_4\right)\left(1-2\lambda_3\right)\sqrt{\left(1-\lambda_1\right)
\left((1+\sqrt{x_{\phi\tau}})^2-\lambda_1(1-\sqrt{x_{\phi\tau}})^2\right)}\right]
\nonumber \\
&& \left. \pm
2(1-\sqrt{x_{\phi\tau}})\sqrt{\lambda_2(1-\lambda_3)\lambda_3(1-\lambda_4)
\lambda_4(1-\lambda_1)
\left((1+\sqrt{x_{\phi\tau}})^2-\lambda_1(1-\sqrt{x_{\phi\tau}})^2\right)} \right] \,. \nonumber \\ &&
\end{eqnarray}


\end{document}